\documentclass{emulateapj}
\usepackage{epsf,graphics}

\newcommand{\kms}{kms$^{-1}$}
\newcommand{\dv}{$r^{1/4}\,$}

\newcommand{\mbh}{M$_{\rm BH}$}
\newcommand{\lb}{L$_{\rm B}$}

\received{10 April 2007}
\accepted{2 June 2007}
\slugcomment{}
\shorttitle{Cosmic evolution of the M$_{\rm BH}-L_{\rm B}$ relation}
\shortauthors{Treu et al.}

\begin{document}

\title{Cosmic Evolution of Black Holes and Spheroids. II: Scaling Relations at $z=0.36$}

\author{Tommaso Treu\altaffilmark{1,2}, Jong-Hak Woo\altaffilmark{1},
Matthew A. Malkan\altaffilmark{3}, Roger D. Blandford\altaffilmark{4}}

\altaffiltext{1}{Department of Physics, University of California,
Santa Barbara, CA 93106-9530; tt@physics.ucsb.edu,
woo@physics.ucsb.edu} 
\altaffiltext{2}{Alfred P. Sloan Research Fellow}
\altaffiltext{3}{Department of Physics and
Astronomy, University of California at Los Angeles, CA 90095,
malkan@astro.ucla.edu} \altaffiltext{4}{Kavli Institute for Particle
Astrophysics and Cosmology, Stanford, CA, rdb3@stanford.edu}

\begin{abstract}
We use high resolution images obtained with the Advanced Camera for
Surveys on board the Hubble Space Telescope to determine morphology,
nuclear luminosity and structural parameters of the spheroidal
component for a sample of 20 Seyfert galaxies at $z=0.36$. We combine
these measurements with spectroscopic information from the Keck
Telescope (paper I) to determine the black hole mass - spheroid
luminosity relation (M$_{\rm BH}$-L$_{\rm B}$), the Fundamental Plane
(FP) of the host galaxies and the black hole mass - spheroid velocity
dispersion relation (\mbh-$\sigma$). The FP is consistent with that of
inactive spheroids at comparable redshifts. Assuming pure luminosity
evolution, we find that the host spheroids had smaller luminosity and
stellar velocity dispersion than today for a fixed \mbh. The offsets
correspond to $\Delta \log L_{\rm B,0}=0.40\pm0.11\pm0.15$ ($\Delta
\log M_{\rm BH} = 0.51\pm0.14\pm0.19$) and $\Delta \log \sigma =
0.13\pm0.03\pm0.05$ ($\Delta \log M_{\rm BH} = 0.54\pm0.12\pm0.21$),
respectively for the \mbh-L and \mbh-$\sigma$ relations (the double
error bars indicate random and systematic uncertainties). A detailed
analysis of known systematic errors and selection effects shows that
they cannot account for the observed offset. We conclude that the data
are inconsistent with pure luminosity evolution and the existence of
universal and tight scaling relations. In order to obey the three
local scaling relations by $z=0$ -- assuming no significant black hole
growth -- the distant spheroids have to grow their stellar mass by
approximately 60\% ($\Delta \log M_{\rm sph}=0.20\pm0.14$) in the next
4 billion years, while preserving their size and holding their stellar
mass to light ratio approximately constant. The measured evolution can
be expressed as $M_{\rm BH} / M_{\rm sph} \propto (1+z)^{1.5\pm1.0}$,
consistent with black holes of a few $10^8$ M$_{\odot}$ completing
their growth before their host galaxies. Based on the disturbed
morphologies of a fraction of the sample (6/20) we suggest collisional
mergers with disk-dominated systems as the physical mechanism driving
the evolution.

\end{abstract}


\section{Introduction}

In the local Universe, most galactic nuclei harbor a supermassive
black hole \citep[e.g.,][]{K+R95}. The mass of the black hole correlates with
global properties of the host, such as the velocity dispersion and
luminosity of the spheroidal (or bulge) component, indicating a
connection between nuclear activity and galaxy formation and evolution
\citep{Mag++98,F+M00,Geb++00,Geb++01,M+H03,H+R04,NFD06,G+D07}. Understanding 
the origin of this relation is a major challenge for cosmological
models and is believed to hold the key to solving several
astrophysical problems such as the role of feedback from nuclear
activity in suppressing star formation in massive galaxies
\cite[e.g.,][]{Gra++04,DSH05,Cro++06b,Mal++06,C+O07}.  

In the standard cosmological scenario, spheroids grow by mergers of
smaller galaxies while black holes grow by accreting surrounding
matter. Depending on the relative timing of the two processes, the
scaling relations between black hole mass and spheroid luminosity and
velocity dispersion (hereafter \mbh-L -- or \mbh-L$_{\rm B}$ to
specify the blue band -- and \mbh-$\sigma$ relations, respectively)
could also evolve with cosmic time. For example -- if the spheroid
evolved passively due to aging of stellar evolutions, changing L but
not $\sigma$, while the black hole grows by accretion -- we would
expect evolution in both the M-L and the M-$\sigma$ relations, with
the latter evolving more slowly than the former. By contrast, if black
holes completed their growth first and the dominant mode of growth now
is the transformation of stellar disks into spheroids, the two
relations would evolve in the opposite sense \citep{Cro06}. The
tightness of the local relationships has been interpreted as evidence
for feedback that synchronizes the relative growth. In this context,
the tight relationships would be naturally reproduced if galaxies and
black holes moved up the M-$\sigma$ relation during merging events, so
that the correlation would appear not to evolve with redshift
\citep{H+K00}.

Detailed theoretical predictions are extremely difficult due to the
daunting range of scales -- ranging from the Mpc scale halo to the pc
scale dynamical sphere of influence of the black hole, to the $\mu$pc
scale of the accretion disk -- and physical processes involved--
radiative transfer, heating and cooling, accretion, just to name a
few. In spite of the challenge, numerous groups have been able to
develop models that are increasingly more successful at reproducing a
variety of observations
\citep[e.g.][]{Gra++04,M+K05,Hop++06c,Cro++06b,Mal++06,D+B07,C+O07}. However,
to this date, the evolution of scaling laws remains very uncertain and
sensitive to the schemes and approximations adopted to deal with the
complex physics (compare for example the recent works by
\citealt{Hop++07} and \citealt{Rob++06a}).

Accurate empirical measurements are needed to discriminate between
scenarios, and provide input on the relative importance of various
physical phenomena as well as on the accuracy of approximations.  With
this goal in mind, a number of groups have started observational
programs to trace the evolution of scaling laws over cosmic time
\citep{Shi++03,TMB04,Wal++04a,A+S05b,Woo++06,McL++06a,Pen++06qsoa,Pen++06qsob,Sal++06}.
However, observers face two fundamental limitations. On the one hand,
-- since the sphere of influence of supermassive black holes is too
small to be resolved at cosmological distances with present technology
-- black hole mass estimates can only be obtained for active
galaxies. Typically, this involves broad line AGN and the dynamics of
the broad line region, with consequent loss of information about the
host galaxy which is swamped by nuclear light. On the other hand -- at
a deeper level -- the evolution of the scaling laws, depends on the
interplay of at least four physical processes: i) black hole
accretion; ii) evolution of the stellar populations; iii) dynamical
evolution of the spheroid, e.g. through mergers; iv) black hole
feedback on star formation. For this reason, even when a scaling law
can be measured as a function of redshift, the interpretation is often
times ambiguous. For example, as discussed by \citet{Pen++06qsob}, how
much of the evolution of the M-L relation is due to evolution in the
spheroid mass and how much is due to evolution of the stellar
populations?

In this paper we address the two fundamental limitations by adopting
the following strategy. First, as in our pilot study \citep{TMB04} and
in the first paper of this series \citep[hereafter paper I]{Woo++06},
we focus on relatively moderate redshift ($z\sim0.36$) and luminosity
(monochromatic luminosity at 5100\AA\, $\sim 10^{44}$ erg s$^{-1}$)
AGN. Although the lookback time is considerably smaller than that of
the most distant quasars studied by other groups, this choice allows
us to determine the host galaxy properties with considerably smaller
uncertainties. Second, we concentrate on a relatively small sample and
measure several independent properties of the host galaxies. Paper I
reported stellar velocity dispersion measurements based on Keck
spectroscopy. This paper presents host spheroid luminosity and size
measurements based on HST-ACS imaging. We use this combined dataset to
study at the same time and for the same sample the \mbh-L and
M-$\sigma$ relations and the Fundamental Plane of host spheroids. The
combination of these diagnostics -- which can be thought as
projections of a more fundamental manifold \citep[e.g.,][]{Hop++07} --
allows us to disentangle stellar mass growth, stellar population
evolution and black hole growth, and helps to identify the processes
at work \citep[see also][]{C+V01,NLC03,BMQ06}.

The paper is organized as follows. In \S~\ref{sec:data}, we summarize
the properties of the sample, describe the HST-ACS observations, and
present our surface photometry. We also derive black hole masses using
nuclear luminosities as measured from HST images and the new
calibration of the broad line region size - nuclear luminosity
relation \citep{Ben++06a}, together with H$\beta$ line widths from
paper I. To construct a suitable local comparison sample, we use Sloan
Digital Sky Survey images to derive new measurements of the spheroid
luminosity of a sample of local Seyferts with \mbh\, available from
reverberation mapping \citep{Pet++04,Ben++06b}.  \S~\ref{sec:results}
describes our main results, i.e. the Fundamental Plane, \mbh-L$_{\rm
B}$, and \mbh-$\sigma$ relation of the distant Seyferts. Detailed
estimates of systematic errors and selection effects are given in
\S~\ref{sec:sys}. \S~\ref{sec:mani} analyzes the three scaling
relations under the assumption that the distant Seyferts will evolve
to match the local relations and derives constraints on the evolution
of stellar mass, size and stellar populations of the host spheroids as
a function of black hole growth. The results are discussed and
compared with the literature in \S~\ref{sec:disc}, and summarized in
\S~\ref{sec:sum}.

Throughout this paper magnitudes are given in the AB scale. We assume
a concordance cosmology with matter and dark energy density
$\Omega_m=0.3$, $\Omega_{\Lambda}=0.7$, and Hubble constant H$_0$=70
kms$^{-1}$Mpc$^{-1}$.

\begin{figure*}[t]
\begin{center}
\plotone{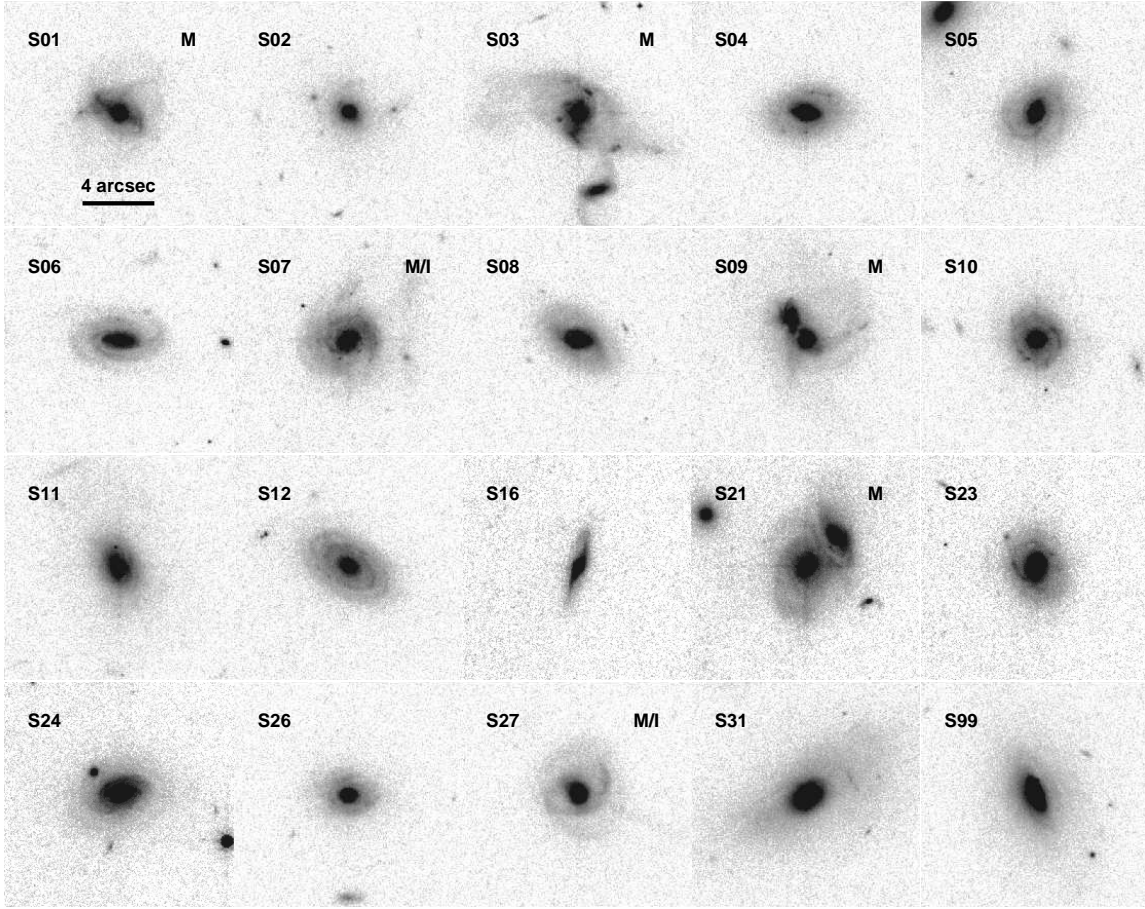}
\end{center}
\figcaption{
Postage stamp images of the 20 Seyfert galaxies in the
sample. Each postage stamp is 12 arcseconds on a side,
i.e. approximately 60kpc. Merging (Interacting) objects are identified
by the label M (I). The centers of galaxies S11, S31 and S99 are
obscured by dust lanes, which prevented accurate surface photometry of
the nucleus and spheroid.
\label{fig:montage}}
\end{figure*}

\section{Data}
\label{sec:data}

\subsection{Sample selection and observations}

The sample discussed in this paper coincides with that discussed in
paper I of this series, with minor exceptions. Two of the objects (S16
and S31) studied in this paper do not have Keck spectroscopy due to
the unfavorable weather conditions at Keck when follow-up observations
were planned.  One of the objects with deep Keck spectroscopy (S28)
has no ACS images because it was observed after the schedule of the
Hubble program had been completed. As in our previous paper, except
for S99 that was selected before SDSS data release 1 (DR1)
\citep{TMB04}, the objects were selected from the SDSS archive, based
on the following criteria: i) spatially resolved in the SDSS images;
ii) redshift between 0.35 and 0.37; iii) H$\beta$ equivalent width and
Gaussian width greater than 5 \AA\, in the rest frame. Most of the
objects were selected from DR-1, and the remaining ones (S21,S23, S26,
S27) were selected from DR-2. After the initial selection based on
these criteria, the SDSS spectra of the objects observable from Keck
were visually inspected by two of us (TT and MAM) to check line
identification. Objects showing strong Fe II nuclear emission (the
main obstacle to velocity dispersion measurement) were eliminated from
the sample. For example, for the run of September 2003, out of 33
observable objects, 8 were rejected on the basis of visual inspection
(the mean and rms $r'$ magnitude of the selected and rejected samples
are consistent, 18.76 and 0.42 vs. 18.98 and 0.63, respectively). Of
the remaining 25 objects, 12 were observed during the run, based on
observability. The fraction of rejected is similar for following runs,
although the total number of objects increases significantly in
subsequent data releases. No color selection was imposed, although a
post-facto analysis shows that the SDSS colors of the sample are
intermediate between those of a quasar and those of an old stellar
population, consistent with the comparable fraction of nuclear and
stellar light inferred from the HST images in the rest of this
paper. Coordinates, redshifts and other basic properties of the sample
are given in Table~\ref{tab:sample}.  

The sample was observed using the Wide Field Camera of the ACS on
board HST between August and December of 2004 as part of General
Observer program 10216 (PI: Treu). Each object was observed for one
orbit, split in four exposures dithered by semi-integer pixel shifts
to ensure cosmic ray and defect removal as well as to improve sampling
of the point spread function (PSF). The total exposure times range
between 2148s and 2360s. Filter F775W ($i'$) was chosen so as to
sample the region corresponding to the rest frame 5100 \AA\, in order
to estimate the size of the broad line region, and to avoid
contamination from the broad emission lines H$\beta$ and
H$\alpha$. This filter choice also provides spheroid luminosity redward
of the 4000 \AA\, break, and it is thus appropriate to infer spheroid
structural parameters and rest frame B and V luminosity with minimal
uncertainty due to filter transformations. Postage stamp images of the
targets are shown in Figure~\ref{fig:montage}.

\subsection{Reduction and analysis}

The following observables are needed to investigate our science
questions: i) spheroid luminosity; ii) nuclear luminosity, to estimate
the size of broad line region, and hence the black hole mass; iii)
effective radius and effective surface brightness of the spheroid, to
construct the Fundamental Plane. This section presents the
measurements with an extensive discussion of systematic and random
uncertainties.

\subsubsection{Reduction}

The images were reduced using {\sc multidrizzle} to remove cosmic rays
and defects, correct for distortion and improve sampling of the
PSF. Based on Monte Carlo simulations we adopted a final pixel size of
$0\farcs04$, which provides the best compromise between sampling of
the PSF, signal to noise ratio of the individual pixels, and noise
correlation.

\begin{figure*}[th!]
\begin{center}
\plotone{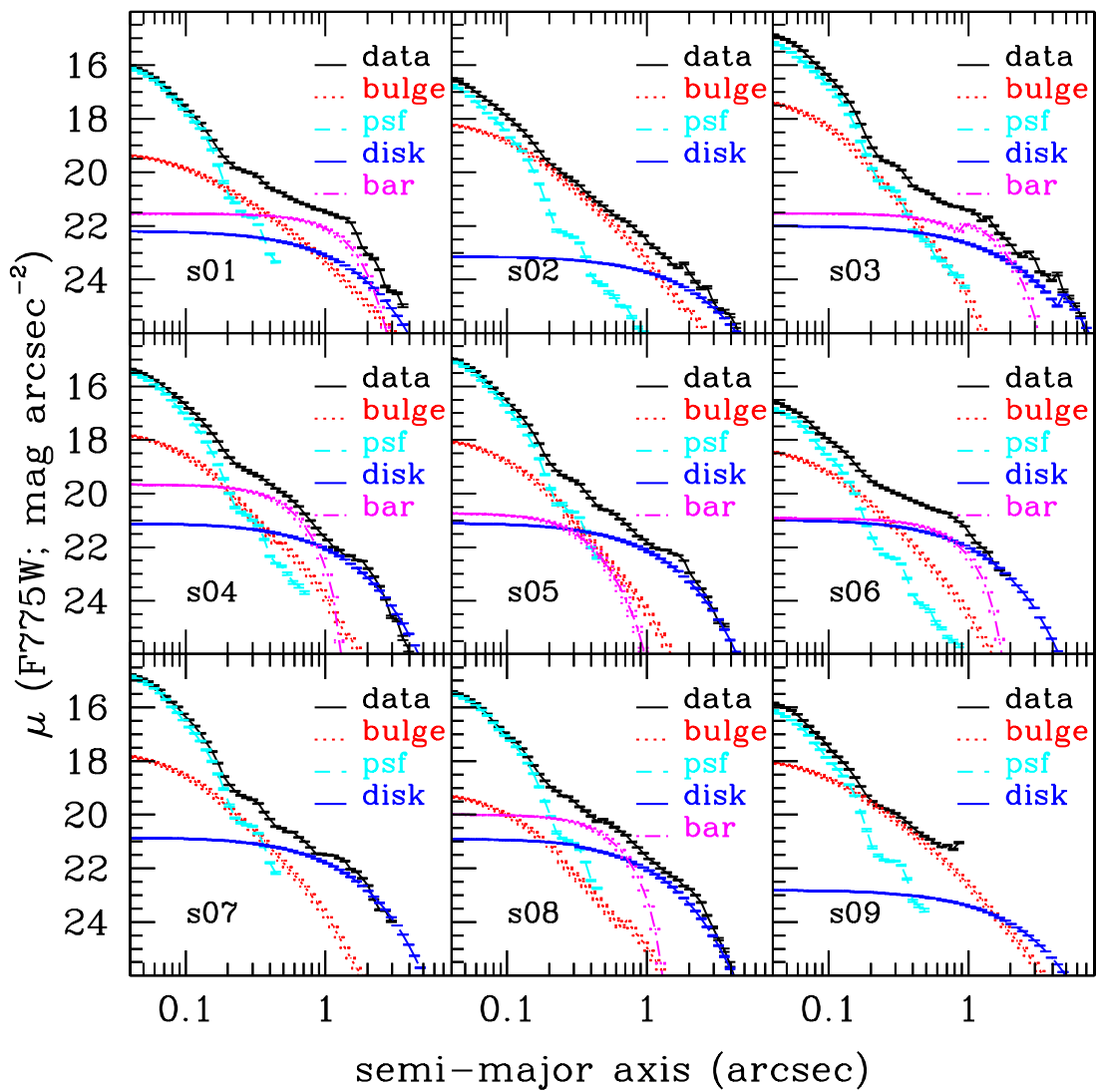}
\end{center}
\figcaption{
Surface photometry I. The surface brightness profile
measured from the data are shown together with that determined from
the two-dimensional model that best fits the data. The surface
brightness profile of each component is shown separately to illustrate
the relative contribution as a function of radius. Note that this plot
is for illustration only, and the fits were performed in twodimensions
as described in Section~\ref{sssec:surf} using cutouts of 20$''$ on a
side. The early truncation of a few profiles (e.g. S09) is only an
artifact of the elliptical isophotes routine used to make the plots,
due to nearby objects. For the measurement, the nearby objects were
modeled and fitted simultaneously in twodimensions.  
\label{fig:profiles1}}
\end{figure*}

\subsubsection{Surface photometry}

\label{sssec:surf}

Surface photometry was derived using the {\sc galfit} software
\citep{Pen++02} to fit two-dimensional models to the data.  Optimizing
a large number of non-linear parameters is a notoriously difficult
problem, so we proceeded in steps, adding one component at a
time. First, we fitted a point source to determine the center of the
system, which is assumed common to all components.  We then added a
spheroid modeled as a \citet{deV48} \dv\, profile, and/or an
exponential disk, if required by the $\chi^2$ statistic.  If needed,
an additional component described by a \citet{Ser68} profile with
index n=0.5 was added to model the bar, identified as an elongated
feature with a change in position angle with respect to the overall
surface brightness distribution (see, e.g., S01 in Figure~1). Bars
were found to be present in 7/20 cases and contribute between 4\% and
16\% of the total light.  Extensive tests were conducted to ensure
that the solution corresponded to the true global minimum of the
$\chi^2$ over the parameter space.  Bright unsaturated stars in the
field of the images were use to create a library of 43 point spread
functions. Each galaxy was fitted with all the PSFs in the library to
find the best fitting one as well as the best fitting parameters.

\begin{figure*}[th!]
\begin{center}
\plotone{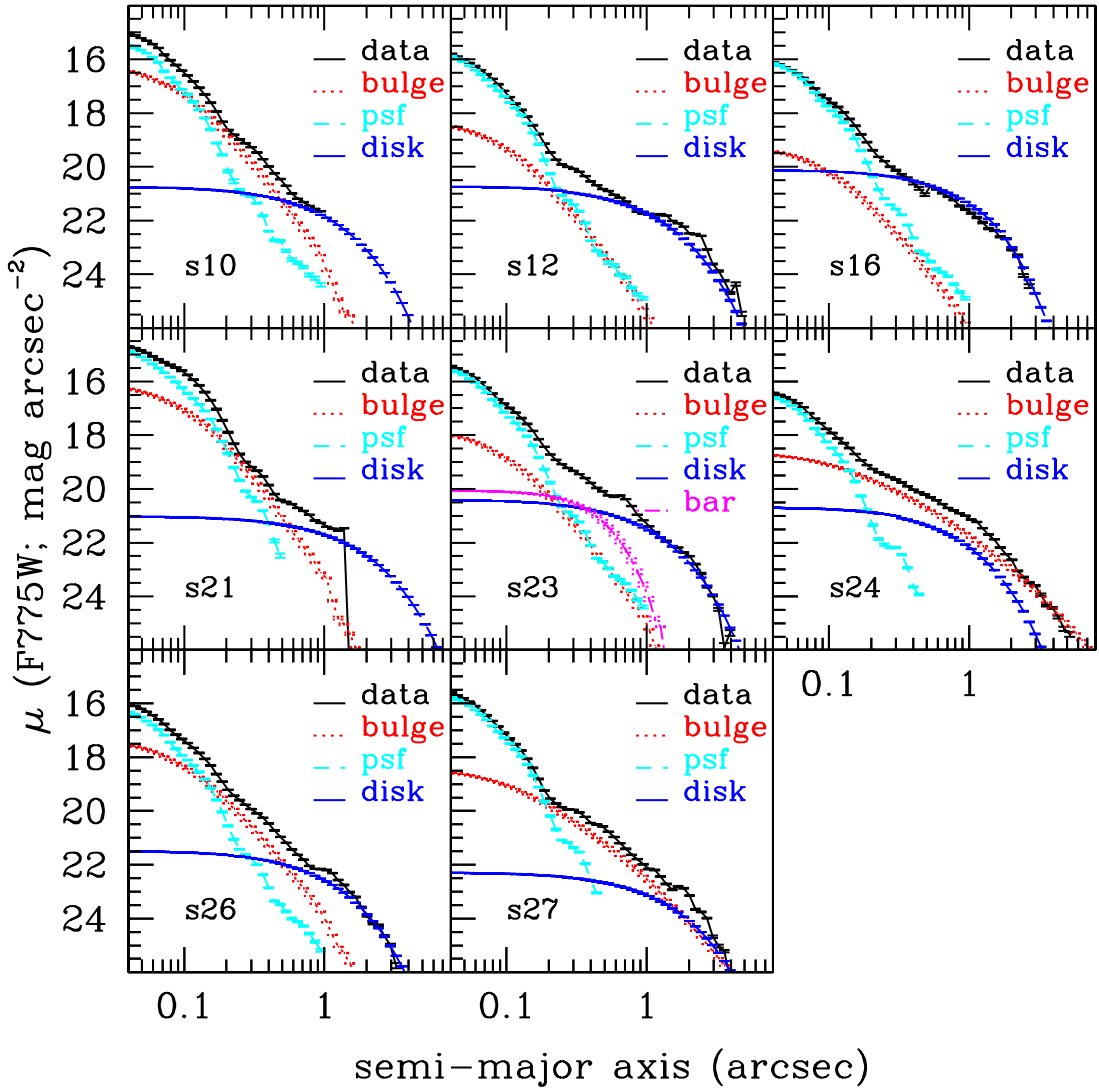}
\end{center}
\figcaption{Surface photometry II. As in Figure~\ref{fig:profiles1} for the rest of the sample.
\label{fig:profiles2}}
\end{figure*}

In three cases (S11, S31, S99), a prominent dust lane prevented us
from determining accurate surface photometry. In two cases (S03 and
S10) no stable solution with a sizable spheroid could be found, as the
spheroidal component tended to become vanishingly small in size. Therefore
we fixed the spheroid half light radius to 2.5 pixels (0.1 arcsec), the
minimum size that could be resolved given our PSF and we considered
the measured luminosity as an upper limit. Similarly, for three objects
(S12, S21, S23) the measured spheroid half light radius is very close to
our resolution limit (i.e. $<$3pixels), and therefore we also consider
their luminosity as an upper limits. Extensive testing shows that the
upper limit to the spheroid luminosity is robust with respect to the
choice of the fixed half light radius, and of the PSF. Twelve out of
twenty cases provided stable best fitting models and hence robust
measurements. For illustration purposes one-dimensional surface
brightness profiles (obtained with {\sc iraf} task {\sc ellipse}) are
shown in Figures~\ref{fig:profiles1} and~\ref{fig:profiles2}.

The formal statistical uncertainties on the spheroid luminosity are
typically 0.05 mags. The rms scatter of the parameters for all the
statistically acceptable PSFs was adopted as the systematic
uncertainty due to PSF modeling. We estimated the total uncertainty --
including systematic errors -- by varying systematically all the
fitting parameters and PSF, finding that the results are typically
stable within 0.2 mags. Spheroid luminosity was then transformed into
rest frame B-band luminosity by correcting for Galactic extinction and
applying K-color corrections, calculated as described in
\citet{Tre++01a}. Errors on extinction and K-color corrections are a
few hundredths of a magnitude. Conservatively, we adopt 0.5 mags
(i.e. 0.2 dex) as the total uncertainty on spheroid luminosity.  This
uncertainty is smaller than the estimated uncertainty on black hole
mass from single epoch measurements $\sim$0.4 dex and therefore
sufficient to meet our goal of constructing the M-L$_{\rm B}$
relationship. Observed and intrinsic spheroid luminosities are given
in Table~\ref{tab:meas}.

\begin{figure}
\begin{center}
\plotone{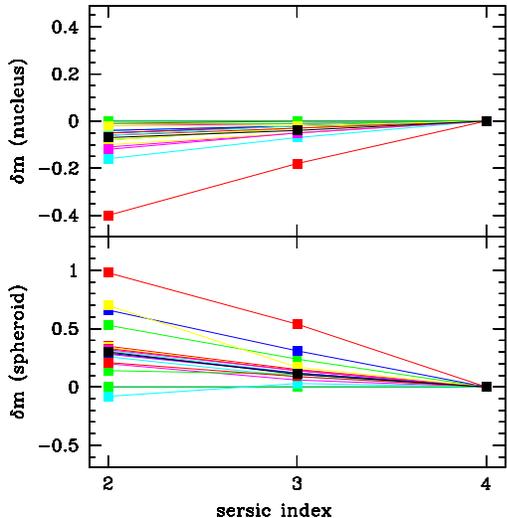}
\end{center}
\figcaption{Systematic effects due to adopted spheroid profile. Changing
the Sersic index of the spheroid from n=4 to n=3 (2), reduces
systematically the spheroid luminosity by 0.15 (0.33) magnitudes, and
increases the nuclear luminosity by 0.02 (0.05) magnitudes.
\label{fig:sersic}}
\end{figure}

Another source of systematic uncertainty is the choice of
the surface brightness profile used to describe the spheroid. Although
\dv\, profiles are the traditional and widely used choice in the
analysis of AGN host galaxies \citep[e.g.][]{Ben++06b}, detailed
studies of nearby bulges show that Sersic profiles with lower Sersic
index (typically 2-3) can provide a better fit. To estimate systematic
errors on the spheroid luminosity associated with our choice of
profile, we repeated our analysis using a Sersic profile with index 2
and 3 (Figure~\ref{fig:sersic}). As expected, we find that the best fit
spheroid luminosity decreases with the Sersic index. Quantitatively,
the spheroid luminosity decreases on average by 0.15 (0.33) mags when
changing the Sersic index from 4 to 3 (2). Similarly, the point source
luminosity increases on average by 0.02 (0.05) magnitudes and the
effective radius of the spheroids decreases by 9\% (11\%) when
changing the Sersic index from 4 to 3 (2). We thus conclude that
adopting a \dv\, profile provides a conservative measurement of the
maximum luminosity of the spheroid and of the minimum nuclear
luminosity (and hence \mbh). We will come back to this point in the
discussion of our results in Section~\ref{ssec:sysL}.  

Other measurements of interest are total magnitude, nuclear luminosity
at rest frame 5100\AA\, (L$_{\rm 5100}$), the combination of effective
radius (R$_{\rm e}$) and effective surface brightness (SB$_{\rm e}$)
that enters the Fundamental Plane FP$_{\rm ph} = \log {\rm R}_{\rm e}$
- 0.32 {\rm SB}$_{\rm e}$ (see~\ref{ssec:FP}) and the fraction of
nuclear light (f$_{\rm nuc}$). The total uncertainties, conservatively
estimated as for the spheroid luminosity, are 20\% on L$_{\rm 5100}$
and on f$_{\rm nuc}$, and 0.1 on FP$_{\rm ph}$. The relevant quantities
are also listed in Table~\ref{tab:meas}.

\subsection{Black hole mass}

As in paper I of this series, black hole masses were derived from the
width of H$\beta$ and the observed nuclear luminosity at 5100 \AA,
using the so-called 'virial' method or empirically calibrated
photo-ionization method \citep{WPM99,Ves02,V+P06}.  Briefly
summarized, the method assumes that the kinematics of the broad line
region trace the gravitational field of the central black hole. The
width of the line provides the velocity scale, while the nuclear
continuum luminosity provides the size via the empirical correlation
observed for the local reverberation mapped sample \citep{WPM99,
Kas++00a,Kas++05,Ben++06a}. The virial coefficient is obtained by
requiring that local AGN hosts and quiescent galaxies obey the same
M-$\sigma$ relation \citep{Onk++04,G+H06ms}. This method has been
shown to provide an estimate of the black hole mass within a factor of
2-3. For this paper we will assume as in the previous paper of the
series a nominal uncertainty of 0.4 dex on black hole mass obtained
with this method (see Peterson 2007, for a recent discussion of the
method). However, since we allow for unknown intrinsic scatter in the
scaling relations when fitting for the intercept
(\S~\ref{sec:results}), adopting a larger random uncertainty on each
individual black hole mass estimate \citep[e.g. 0.6 dex,][]{V+P06} has
a negligible impact in terms of overall uncertainty
(\S~\ref{ssec:random}).

As in paper I, we adopt the second moment of the broad component of
H$\beta$ as our velocity scale since this is more robustly measured
than the alternative FWHM (paper I; see also Peterson et al. 2004;
Collin et al.\ 2006). The velocity scale is measured from Keck spectra
when available and from SDSS spectra in the few cases when Keck
spectra are not available (see paper I for details). As shown in paper
I, our methodology gives unbiased black hole masses when applied to the
local sample of calibrators. A major improvement with respect to paper
I is that we can use high resolution Hubble images to measure the
nuclear luminosity and infer broad line region size. In practice we
use the following equation to estimate black hole masses, obtained
combining the most recent calibrations of the size luminosity relation
and virial coefficients \citep{Onk++04,Ben++06b}.

\begin{equation}
\log M_{\rm BH} = 8.58 + 2 \log \left(\frac{\sigma_{\rm H\beta}}{3000 {\rm kms}^{-1}}\right) + 0.518 \log \left( \frac{L_{\rm 5100}}{10^{44} {\rm erg s}^{-1}}\right),
\label{eq:MBH} 
\end{equation}

\noindent
where $\sigma_{\rm H\beta}$ is the second moment of the broad H$\beta$
line profile. We note that for the sample in common with paper I, the
use of nuclear luminosities and this new calibration implies black
hole masses smaller by 0.09 dex on average, consistent with our
estimate of the magnitude of this systematic error in paper I. Black
hole masses are listed in Table~\ref{tab:meas}.

\subsection{Local comparison samples}

To study evolutionary effects we consider two comparison samples to
define the local M-L$_{\rm B}$ relation. First we consider the sample
of quiescent galaxies collected by \citet{M+H03}. The sample consists
of galaxies with black hole mass determined from spatially resolved
kinematics and with spheroid luminosity determined from
two-dimensional surface photometry. The second sample is that of local
Seyfert galaxies with black hole mass measured via reverberation
mapping discussed in \citet{Pet++04} and \cite{Onk++04}. To ensure
self-consistent determination of spheroid luminosity for the local and
distant Seyfert sample we measured the parameters of the local sample
in exactly the same way as we do for distant galaxies. We searched the
SDSS archive for g'-band images of local Seyferts, which provide a
very good match to our distant galaxies in terms of resolution (the
seeing is typically 10 times the HST PSF, but the angular size
distance is typically 10 times smaller) and rest frame
wavelength. SDSS images for nine Seyferts were found in the
archive. The resulting spheroid luminosities for the local Seyfert
samples are listed in Table~\ref{tab:local}, together with redshifts,
black hole masses and velocity dispersions from the literature.

The two samples are complementary in terms of vices and virtues.  The
first sample is larger in size and black hole masses, and spheroid
luminosities are most robustly measured: i) \mbh\, is obtained from
spatially resolved kinematics; ii) the determination of spheroid
luminosity does not suffer from poor resolution or the presence of a
prominent point source in the center. However, the second sample is
the most appropriate for a direct comparison for a variety of reasons:
i) the virial coefficient is unknown, but assuming that it is not
evolving with redshift, the comparison between Seyfert samples is
independent of its numerical value; ii) if the local M-L relation is
not universal, Seyferts may define a different M-L relation than
quiescent early-type galaxies -- perhaps because they are at a
different evolutionary stage. Thus, we conclude that at this stage the
more conservative approach is to consider both samples for the
following analysis and consider the uncertainty on the local relation
as an additional source of systematic errors. More comprehensive
studies of the local M-L relation are needed to eliminate this source
of uncertainty. This is left for future work.

\section{Results}
\label{sec:results}

This section presents the main results of this paper. First in
\S~\ref{ssec:FP} we discuss the Fundamental Plane of the host galaxies
in comparison with that of normal quiescent galaxies, for the
subsample of objects that have both $\sigma$ and structural
parameters. In \S~\ref{ssec:ML} we present the \mbh-\lb\, relation for
distant Seyfert galaxies. Having established that the Fundamental
Plane of the host galaxies is indistinguishable from that of quiescent
galaxies, we adopt the luminosity evolution inferred from FP analysis
\citep{Tre++05a,Tre++05b} to compare with the local \mbh-\lb\,
relation.  In \S~\ref{ssec:MS} we revisit the \mbh-$\sigma$ relation
derived in paper I, using the new improved black hole mass estimates
based on nuclear luminosities determined from HST imaging.

\subsection{Fundamental Plane}
\label{ssec:FP}

In the local Universe early-type galaxies obey the Fundamental Plane,
i.e. an empirical correlation between effective radius, central
velocity dispersion and effective surface brightness of the form:

\begin{equation}
\log R_{\rm e} = \alpha \log \sigma + \beta {\rm SB_{\rm e}} + \gamma,
\label{eq:FP}
\end{equation}

\noindent
with $\alpha=1.25$, $\beta=0.32$ and $\gamma=-9.00$ for the Coma
Cluster in the B(AB) band. The evolution of the FP out to redshift
$z\sim0.4$ is well established for quiescent early-type galaxies
\citep[e.g.,][and references therein] {Tre++05b,Tre++01b}, for AGN
hosts \citep{Woo++04,Woo++05}, and for the spheroidal component of
spiral galaxies with bulge-to-total luminosity ratio greater than 0.2
(MacArthur et al. 2007, in preparation).

\begin{figure}
\begin{center}
\plotone{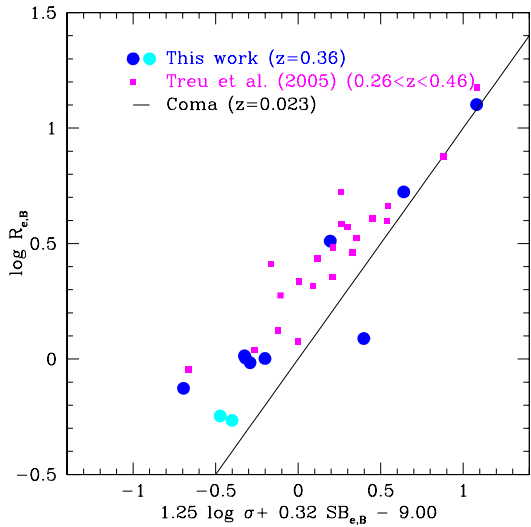}
\end{center}
\figcaption{Edge on view of the Fundamental Plane of the distant
Seyfert galaxies (circles) and of normal quiescent galaxies at
comparable redshift (squares). Seyfert galaxies for which only an upper limit to
the spheroid luminosity is available are plotted as cyan symbols. The
errors on the FP variables are highly correlated, and points are
allowed to move mostly within the plane. The error component
perpendicular to the plane is estimated to be $\sim 0.1$ dex for the
Seyfert sample when projected along the $\log R_{\rm e}$ axis. Most of
the apparent thickness of the FP for quiescent galaxies is due to the
relatively large range in redshifts and luminosity evolution during
the corresponding interval in cosmic time. The local FP of the Coma
Cluster is shown as a solid black line for comparison.
\label{fig:FPB}}
\end{figure}

In Figure~\ref{fig:FPB} we plot the FP parameters of the subset of
distant Seyferts for which structural parameters and stellar velocity
dispersion is available, together with a comparable sample of
quiescent galaxies taken from \citet{Tre++05a,Tre++05b}. The good
agreement between the FP of quiescent and active galaxies gives us
confidence that our surface photometry is not systematically biased by
the presence of a nuclear point source. The offset with respect to the
local relation (solid line) is normally interpreted as evolution of
the stellar populations. Under the assumption of pure luminosity
evolution, luminosity evolves with redshift such that the expected
value at $z=0$ is given by
\begin{equation}
\log L_{\rm B,0} = \log L_{\rm B} - (0.72\pm0.06\pm0.04) z 
\label{eq:lbev}
\end{equation}
\citep{Tre++05b}. Applying the same assumption to the sample of distant
Seyferts at $z=0.36$ implies that: $\log L_{\rm B,0} = \log L_{\rm B}
- 0.26 \pm 0.03$, where random and systematic errors have been added
in quadrature for simplicity.

\subsection{M$_{\rm BH}$-L$_{\rm B}$ Relation}
\label{ssec:ML}

In order to compare with the local relations we need to account for
the fact that the luminosity of a stellar population decreases as it
ages. As our first working hypothesis, in this Section we will present
our results assuming that stellar populations evolved as inferred from
the FP studies under a pure evolution scenario, and use the variable
L$_{\rm B,0}$ obtained using equation~\ref{eq:lbev}. A more general
discussion will be given in \S~\ref{sec:mani}.

\begin{figure}
\begin{center}
\plotone{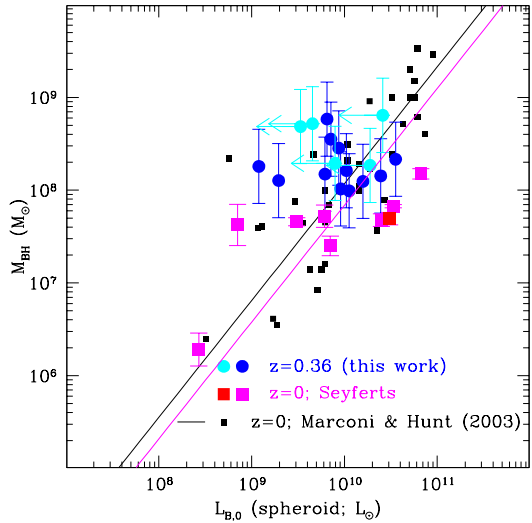}
\end{center}
\figcaption{Black hole mass spheroid luminosity relation. Distant
Seyferts are shown as large circles with error bars. Blue circles
represent measurements, cyan circles with leftward arrows represent
upper limits to the spheroid luminosity. Local Seyferts are shown as
large squares with error bars. Black hole masses are from
\citet{Pet++04} and \citet{Ben++06a}, spheroid luminosities are from
this work (Table~\ref{tab:local}). Magenta squares represent
measurements, the red square represents an upper limit to the spheroid
luminosity. Small black squares represent local quiescent galaxies
from \citep{M+H03}. The black line is the best fits to the
\cite{M+H03} sample. The magenta line is the best fit to the local
Seyferts. For direct comparison the spheroid luminosity of the distant
Seyferts has been evolved to $z=0$ using $\log$L$_{\rm B,0}=\log
L_{\rm B}-0.26$ (see \S~\ref{ssec:ML}).
\label{fig:BHL}}
\end{figure}

\begin{figure}
\begin{center}
\plotone{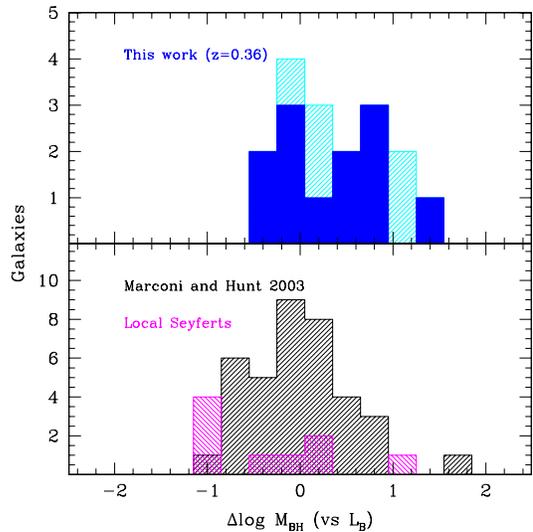}
\end{center}
\figcaption{Distribution of residuals in $\log$\mbh\, with respect to
the fiducial local relation of \citet[solid black line in
Figure~\ref{fig:BHL}]{M+H03}. The Upper panel shows distant Seyferts
(measurements in blue, upper limits in cyan; note that upper limits in
$\log$L$_{\rm B,0}$ correspond to lower limits in $\Delta \log$ \mbh). The
lower panel shows the distribution of residuals for the local samples
of quiescent and Seyfert galaxies.
\label{fig:BHLhisto}}
\end{figure}

The relation between black hole mass and host spheroid luminosity for
our points as well as for local comparison samples is shown in
Figure~\ref{fig:BHL}. The main result is that the Seyfert galaxies at
$z=0.36$ and those at $z=0$ cover approximately the same range in
spheroid luminosity, but the average black hole mass is higher for the
distant Seyferts. The mismatch is exacerbated if one considers that
five of the distant Seyferts measurements are upper limits (while only
one of the local Seyferts is an upper limit). Conservatively we will
generally consider the measured offset as the best estimate of the
offset although it should be kept in mind that it is most likely a
lower limit. Quantitatively, the offset with respect to the fiducial
local relationship (solid line) corresponds to $\Delta L_{\rm B,0} =
0.32\pm0.11\pm0.15$ ($\Delta \log M_{\rm BH} = 0.40\pm0.14\pm0.19$),
considering the intrinsic scatter of the relation to be a free
parameter and then marginalizing over it. Listed systematic errors are
derived as discussed in Section~\ref{sec:sys}. If measured with
respect to the local Seyferts (magenta line), the best estimate of the
offset changes to $\Delta \log M_{\rm BH}=0.63$. This is visualized in
Figure~\ref{fig:BHLhisto} where we plot the distribution of residuals
with respect to the local fiducial relation for the distant and local
samples.

We conclude that pure luminosity evolution is inconsistent with the
observations, if we required that at $z=0$ all galaxies obey the
\mbh-L$_{\rm B}$ relation. We cannot solve this inconsistency by
invoking different luminosity evolution. The stellar populations would
be required to become brighter with time in order to reconcile the
data with the local relationship. This can be ruled out on physical
grounds. Therefore, taking this result at face value, we have to
conclude that either not all spheroids obey the M-L relations, or that
a significant amount of new stars have been added to the spheroids
since $z=0.36$. The interpretation of this result will be discussed in
Sections~\ref{sec:mani} and~\ref{sec:disc}, after we conclude
presenting the evidence and discussing systematic errors and selection
effects in the remainder of this section.

\subsection{M$_{\rm BH}$-$\sigma$ Relation}
\label{ssec:MS}

\begin{figure}
\begin{center}
\plotone{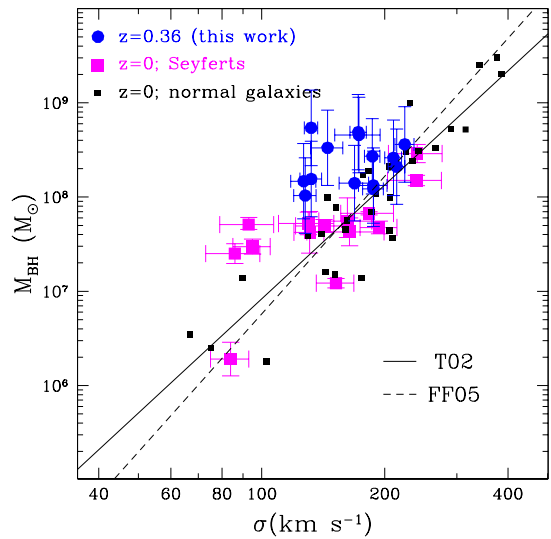}
\end{center}
\figcaption{Black hole mass velocity dispersion relation. The large blue 
circles represent distant Seyferts. Black hole masses are from this
work, velocity dispersions are from paper I. The large magenta squares
represent local Seyferts, the small black squares represent local
quiescent galaxies. The solid and dashed lines are the best fit
relations from \citet{Trem++02} and \citet{F+F05}. The local Seyferts
obey the same \mbh-$\sigma$ relation as local quiescent galaxies by
construction, see \citet{Onk++04} for discussion.
\label{fig:Ms}}
\end{figure}

Figure~\ref{fig:Ms} shows the \mbh-$\sigma$ relation for the distant
Seyferts as well as the local comparison samples. The samples and
symbols are the same as in Figure~\ref{fig:BHL} with few minor
exceptions: i) only the 14 distant Seyferts with available stellar
velocity dispersion (from paper I) are plotted, including S28 and S99
for which HST photometry is not available. ii) local relations are
from \citet{Trem++02} and \citet{F+F05}; iii) Local Seyferts obey the
same relation as quiescent galaxies by construction, as this is the
constraint used to derive the virial coefficient \citep{Onk++04}, so
there is no need to show a separate local relation for Seyferts.  The
only substantial difference with respect to Figure~8 in paper I is
that black hole masses have been recalculated based on HST photometry.

The \mbh-$\sigma$ relation shows the same basic result as the
\mbh-L$_{\rm B}$ relation. At fixed host galaxy properties, distant
Seyferts appear to host larger black hole mass than local
ones. Quantitatively, the offset corresponds to $\Delta \log \sigma =
0.13\pm0.03\pm0.05$ (i.e. $\Delta \log M_{\rm
BH}=0.54\pm0.12\pm0.21$).  Note that offset is slightly reduced (0.08
in $\Delta \log M_{\rm BH}$) with respect to paper I and the
systematic error is smaller due to the improved estimate of black hole
masses that avoid stellar contamination. As shown in
Figure~\ref{fig:Ms} and discussed in paper I, adopting the local
relation from \citet{Trem++02} or \citet{F+F05} changes the offset by
much less than the error bars because the two local relations are very
well matched in the range of black hole mass and stellar velocity
dispersion covered by our sample.

\section{Systematics and selection effects}
\label{sec:sys}

This section is devoted to understanding and estimating systematic
errors. First, in \S~\ref{ssec:sysmbh} and \S~\ref{ssec:sysL}, we list
potential sources of systematic error and estimate as accurately as
possible the associated uncertainty. For brevity we will not repeat
the analysis of systematics already discussed in paper I, unless when
substantial changes/improvements are introduced. \S~\ref{ssec:random}
derives the uncertainty on our estimate of the random errors due to
our assumed accuracy of single-epoch black hole mass
determinations. Then, in \S~\ref{ssec:sel}, we introduce a Monte Carlo
scheme to simulate unknown selection effects and estimate potential
biases. \S~\ref{ssec:syssum} gives a very short summary of this
Section for the impatient reader.

\subsection{Is \mbh\, overestimated?}
\label{ssec:sysmbh}

Understanding the \mbh\, estimate is clearly important. While velocity
dispersion and spheroid photometry are measured independently and
present mutually consistent results (as supported by the FP relation,
which does not involve \mbh), only one measurement of \mbh\, is
available and that determines the evolution of both relationships.
Our measurement of \mbh\, depends on two observables (velocity scale and
nuclear luminosity), on the empirically calibrated relationship
between nuclear luminosity and broad line region size and on the
choice of the virial coefficient.  We now go through each source of
error and estimate the systematic error that they may introduce for a
sample of our size.

\subsubsection{H$\beta$ line width and velocity scale}

In paper I we demonstrated that our measurement of the velocity scale
from the second moment of H$\beta$ in single epoch spectra is
unbiased. This means that -- when applied to the sample of local
calibrators -- our estimator yields the same \mbh\, as obtained by the
original studies based on the second moment of H$\beta$ measured from
the variable part of the spectra. Recently, \citet{Pet07} suggested
that single epoch spectra may overestimate the line width by
$\sim$20\%, corresponding to an error of $\sim$0.15 dex on black hole
mass, if the same virial coefficient is assumed. This effect is likely
to depend on the setup of the experiment and on the variability
pattern. However, to be conservative, we will consider this as a
global uncertainty to the zero point of our measured black hole
mass. \citet{Woo++07} studied the time variability of the width of
H$\beta$ for a subset of our objects. They found that the
r.m.s. variations of the second moment of H$\beta$ are less than 14\%
(i.e. $\sim0.1$ dex in \mbh). For our sample of 14-17 objects this has
therefore negligible effect on the average with respect to the
aforementioned source of error. Our best estimate of overall
systematic uncertainty due to line width measurement is 0.15 dex.

To check for systematic effects due to systematic change of the line
shape with redshift or sample properties such as Eddington Ratio
\citep[e.g.,][]{Col++06} we also computed black hole masses using the
full width half maximum instead of the second moment of the line as
velocity scale. After removing the narrow component of H$\beta$ using
[\ion{O}{3}] as a template as described in
\citet{TMB04} and \citet{Woo++06}, we measured the FWHM (McGill et al. 2007, 
in preparation) and derived black hole mass estimates Using Eq.~1 in
\citet{N+T07}. This alternative estimate of black hole mass 
(M$_{\rm BH,NT}$) agrees very well with that adopted in this paper,
with an average offset of $\langle \log {\rm M_{\rm BH}/M_{\rm BH,NT}}
\rangle=-0.02\pm0.05$ dex, i.e. consistent with no offset. 

\subsubsection{Nuclear luminosity and size luminosity relation}

The new calibration of the size-luminosity relation \citep{Ben++06b},
and our HST based nuclear luminosities are a substantial improvement
with respect to paper I, allowing us to avoid issues related to host
galaxy contamination. The dominant source of error is now the
intrinsic scatter of the size luminosity relation (30\%) corresponding
to $\sim0.1$ dex in \mbh\, per individual object. For our sample size,
this translates into a negligible source of error in the mean. As far
as the overall calibration of the relation is concerned, for the
luminosity range of our sample, the 68\% range on the slope and
intercept of the relationship translate into an overall uncertainty in
size of order 0.02 dex. We thus conclude that the uncertainty due to
the (known) scatter of the size luminosity relation is negligible.

\subsubsection{Virial coefficient and zero point of the local relation} 

The final source of error is related to the virial coefficient needed
to convert size and velocity scales into mass and thus to the zero
point of the local relation. As discussed in detail in paper I, we
adopt the virial coefficient determined by \citet{Onk++04} requiring
that the M-$\sigma$ relation be the same for local quiescent and
active galaxies. The uncertainty on the average of the virial
coefficient is 34\%, thus corresponding to an uncertainty of 0.13 dex
on the calibration of the black hole mass. However -- as discussed in
paper I -- assuming that the virial coefficient does not evolve with
redshift, this factor cancels out between the local and distant
samples and therefore is irrelevant.  Incidentally, we note that our
results can also be interpreted as cosmic evolution of the virial
coefficient, if one is willing to drop this assumption. We consider
this an unlikely explanation, as it would require the geometry or
kinematics of AGN to evolve with cosmic time, but unfortunately it
will not be possible to discard it until direct measurements of the
virial coefficient can be obtained in some other way.

The situation is slightly more complex for the \mbh-L relation. Once
the virial coefficient is fixed there is no more freedom.  Therefore
the difference between the intercept of the relation for local active
and quiescent (0.23 dex in \mbh) is an additional source of
uncertainty. Larger samples of galaxies with well determined black
hole mass and spheroid luminosity are needed to overcome this
limitation. For the purpose of this paper we consider the difference
as an additional source of uncertainty. To produce a single
measurement, we weight the two samples equally (for reasons discussed
in \S~3) and take the average and semi-difference of the two intercepts
as best estimate of the local average and systematic uncertainty on
the zero point ($\sim0.12$ dex).

\subsection{Is L$_{\rm B}$ underestimated?}
\label{ssec:sysL}

The main sources of uncertainty that could affect the spheroid
luminosity of the sample as a whole are systematic errors in
K-corrections, in the adopted luminosity evolution, and in the choice
of the surface brightness profile for the bulge. The uncertainty on
the K-correction is negligible, at most 0.02 dex, as estimated using a
range of stellar population models to compute the transformation from
observed F775W to rest frame B. The luminosity evolution as measured
from the Fundamental Plane \citep{Tre++05a} carries an overall
uncertainty of 0.03 dex. Even assuming that this translates to a shift
of the whole sample, this is still a negligible source of error. As
discussed in Section~\ref{sssec:surf} adopting a Sersic profile
instead of a \dv\, profile tends to systematically reduce the bulge
luminosity while increasing the point source luminosity, and therefore
black hole mass, thus moving points further away from the local M-L
relation of quiescent galaxies. In contrast, since we used the same
identical technique for the local Seyferts, adopting a Sersic profile
would not alter the observed offset.  We conclude that the measured
offset is not overestimated by an appreciable amount due to known
systematic errors on spheroid luminosity determination.

\subsection{Are random errors underestimated?}
\label{ssec:random}

Finally, the uncertainty of black hole mass estimates from single
epoch data is assumed to be 0.4 dex \citep{Ves02}. This represents all
sources of random error that contribute to the scatter in
Eq.~\ref{eq:MBH}. After quantifying sources of systematic errors on
the mean of Eq~\ref{eq:MBH}, we now consider whether the random error
associated with our measurement of the offset of the scaling relations
could be underestimated. To test this, we repeat the analysis assuming
that the random error on each individual \mbh\, estimate is 0.6 dex
\citep{V+P06}. The inferred random error on $\Delta \log {\rm M}_{\rm
BH}$ increases by 0.03 dex, e.g. from 0.14 dex to 0.17 dex for the
\mbh-L relation.  Note that our analysis includes a nuisance parameter
to account for the unknown intrinsic scatter of scaling relations and
therefore the estimated errors are slightly different than what would
be naively derived considering only errors on the y axis (0.4 /
$\sqrt{N-1}$ = 0.4/4 = 0.1 dex vs 0.6/$\sqrt{N-1}$=0.15). This is a
negligible error on the error, considering that the systematic term is
dominant. The total error, defined as the quadratic sum of the random
and systematic uncertainties, would only change by 0.01 dex.

\subsection{Selection bias}
\label{ssec:sel}

In this section we estimate possible bias due to selection effects.
Our galaxies are selected based on their nuclear properties (having a
broad line AGN). Therefore, intrinsic scatter and observational errors
in the scaling relations could conceivably lead us to favor large
black hole masses and therefore overestimate evolution\footnote{This
bias is similar to Malmquist bias for luminosity selected samples of
standard candles}.

Our present sample covers approximately a decade in black hole mass
($10^8 -10^9 M_{\odot}$). The upper limit is naturally expected
because of the steep drop of the black hole mass function. Is the
lower limit in mass a result of some unknown selection effect?  Our
nominal selection limits on line width and flux are small enough that
we would have been able to select objects with black hole masses as
low as 10$^7$ M$_{\odot}$, as verified by running our measurement
tools on the entire SDSS-DR5 spectroscopic database at this
redshift. However the objects with \mbh\, well below 10$^8 M_{\odot}$
are a small fraction of the total.  This maybe be due to an intrinsic
drop in the black hole mass function, to a selection effect in the
SDSS spectroscopic sample or to a combination of the two. The decline
in the number of objects at $z=0.36$ with \mbh\, well below 10$^{8}$
is also seen in Figure~1 of \citet{N+T07}, where most of the points
lie above the group identified as having \mbh\, between 10$^{7.5}$ and
10$^{7.8}$ M$_{\odot}$. Estimating the black hole mass function is
beyond the scope of this work \cite[e.g.,][]{Ber++06b,G+H07}. However,
to understand the implications of this selection effect on our
measurement, we can take a conservative approach and model this drop
as a sharp selection in \mbh. In the following discussion we assume
for reference that our implicit selection function $\log {\rm M_{\rm
BH}/M_{\odot}}>7.9$ (the results are unchanged if the limit is chosen
to be 7.8 or 8.0). We then use Monte Carlo simulations to determine
the amount of bias on the offset of the intercept of the scaling
relations introduced by the implicit selection process. The scaling
relations are populated according to the velocity dispersion function
and spheroid luminosity function determined by \citet{Shet++03} and
\citet{Dri++07} respectively, and errors on both axis are taken into
account. We assume that the intrinsic scatter of the relation is
smaller than the measurement errors (0.4-0.5 dex) which is consistent
with local estimates. Note that in this scheme all the selection
procedure is modeled a single hard threshold in black hole mass and we
only work in the scaling relation variables. This simple scheme allows
us to bypass all the assumptions that would be needed to simulate from
first principles the observational selection effects which are a
complex combination of total flux, broad line flux, broad line width
(at both ends), bulge luminosity and stellar velocity dispersion.

\begin{figure}
\begin{center}
\plotone{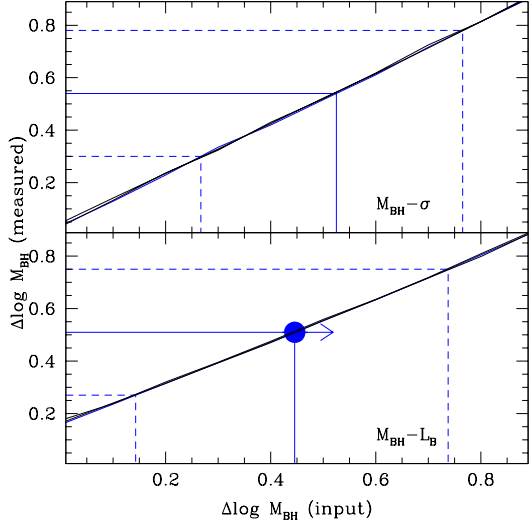}
\end{center}
\figcaption{Estimate of possible bias due to selection effects. The effects 
on an hypothetical selection $\log(M_{\rm BH}/M_{\odot})>7.9$ are
calculated via Monte Carlo simulations (see \S~\ref{ssec:sel} for
details). The measured offset is shown as a function of the simulated
input offset in \mbh\, with respect to the local relation. {\bf Upper
panel: results for the \mbh-$\sigma$ relation}. The bias is negligible
due to the small error on $\sigma$ (compared to that on
\mbh) and the flatness of the velocity dispersion function in the
range of interest. The measured offset and the corresponding input
offset are identified by the solid blue lines. Dashed lines identify
error bars obtained by summing in quadrature random and systematic
errors. {\bf Lower panel: results for the
\mbh-L$_{\rm B}$ relation}. The bias is somewhat more significant due to the
larger error on L$_{\rm B}$ and to the steeper faint end of the
spheroid luminosity function. The solid blue line and large filled
circle identify the lower limit to the offset as measured with respect
to the average of local quiescent and active galaxies.  Dashed lines
represent error bars as in the upper panel. \label{fig:bias}}
\end{figure}

The results of the simulations are shown in Figure~\ref{fig:bias}.
The curves show the recovered (``measured'') offset as a function of
the input offset $\Delta \log$ \mbh. The upper panel shows the results
for the \mbh-$\sigma$ relation, the lower panel shows the results for
the \mbh-L relation. The bias is almost negligible for the
\mbh-$\sigma$ relation, while it is at most 0.1 dex for the \mbh-L
relation. The difference between the two relations is due to: i) the
smaller errors on velocity dispersion than those on spheroid
luminosity; ii) the different behavior of the two distributions at the
faint (low velocity) end. The velocity dispersion function peaks in
the range covered by our sample, while the luminosity function of
bulges extends to much smaller luminosities. As a softer cutoff would
imply smaller bias, we conclude that selection effects are responsible
for at most 0.1 dex of the observed offset of the \mbh-L relation, and
that this source of bias is negligible for the \mbh-$\sigma$ relation.

\subsection{Summary of uncertainties and best measurements}
\label{ssec:syssum}

In conclusion, the total systematic error on the evolution of
\mbh-$\sigma$ relation is 0.21 dex in \mbh\, (i.e. slightly reduced with
respect to paper I due to the improved nuclear luminosity estimates),
and 0.19 on the \mbh-L relation, dominated by the uncertainty on the
black hole mass and on the uncertainty on the local relation.

Although we caution the reader to keep in mind all the caveats
discussed above, we now condense all the information discussed in this
section in seven numbers. Our best estimates of the offset of the
relations -- without accounting for selection effects are:

\begin{itemize}
\item \mbh-L: $\Delta \log M_{\rm BH} = 0.51\pm0.14\pm0.19$
\item \mbh-$\sigma$: $\Delta \log M_{\rm BH} = 0.54\pm0.12\pm0.21$
\end{itemize}

Unknown selection effects could remove as much as 0.1 dex to the
offset of the \mbh-L relation.

\section{Evolution of the Fundamental Manifold of Black Holes and Spheroids}
\label{sec:mani}

In the previous sections we concluded that pure luminosity evolution
is inconsistent with the observations and the requirement that the
distant Seyferts lie on the local relations. In this section, we
explore the evolutionary constraints that can be obtained by examining
the evolution of black holes and spheroids in a four dimensional
parameter space -- with axes given by luminosity, velocity dispersion,
size and black hole mass -- instead that on lower dimension spaces.

In other words, we know that in the local universe, black holes and
their host spheroids lie on the FP, the \mbh-$\sigma$ and \mbh-L
relations -- 3 relationships involving 4 parameters. Thus, by requiring
that our distant galaxies land on the local relationships at $z=0$, we
can derive constraints on the evolution in size, luminosity and mass
of the host galaxies, as a function of black hole growth in the same
time span.

\begin{figure}
\begin{center}
\epsscale{0.9}
\plotone{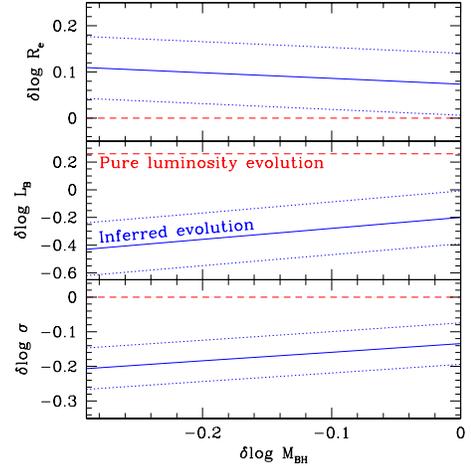}
\end{center}
\figcaption{Inferred evolution of the fundamental manifold parameters
(solid blue lines), as a function of change in black hole mass. Dotted
blue lines represent the 68\% confidence bands. The symbol $\delta$
indicates the difference between the parameter at $z=0.36$ and the
parameter at $z=0$. The red dashed horizontal lines represent pure
luminosity evolution of the host galaxies for reference. If there is
no black hole growth ($\delta \log M_{\rm BH}$=0) the host galaxies of
the distant Seyferts need to increase their velocity dispersion by
$0.13\pm0.06$ dex and increase their luminosity by $0.20\pm0.19$ dex
in the next 4 Gyrs in order to obey simultaneously the local FP,
\mbh-$\sigma$, and \mbh-L$_{\rm B}$ relations, with no significant
change in the effective radius. For comparison, pure luminosity
evolution of the spheroid would predict 0, -0.26 and 0, respectively
(dashed red lines).  This corresponds to a growth of $\sim $60\% of
the spheroid stellar mass and approximately constant stellar
mass-to-light ratio. If there is significant black hole growth
($\delta \log$\mbh $<0$) the increase in $\sigma$ and L$_{\rm B}$ and
the decrease in effective radius must be more pronounced.
\label{fig:manifold}}
\end{figure}

The results of this exercise are shown in
Figure~\ref{fig:manifold}. For simplicity, systematic and random
errors are combined in quadrature in order to compute confidence
bands, and neglecting covariance. The main result is that in order to
satisfy all scaling relations in the local universe the host galaxies
have to grow in luminosity and velocity dispersion, while leaving the
size substantially unchanged.

Consider the no black hole growth scenario (consistent with the low
Eddington ratios, see paper I).  Assuming that M$_{\rm sph}\propto
\sigma^2 R_{\rm e}$, the spheroid mass has to increase by
$0.20\pm0.14$ dex ($\sim 60$\%), by today. In other words,

\begin{equation}
\frac{M_{\rm BH}}{M_{\rm sph}} \propto (1+z)^{1.5\pm1.0}
\label{eq:evz}
\end{equation}

Similarly, the spheroid luminosity has to increase by $0.20\pm0.19$
dex, implying that its average stellar mass to light ratio has to stay
approximately constant (with an uncertainty of approximately 50\%)
over the next 4 Gyrs. This requires an injection of younger stars to
counteract the ageing of the resident population. The effective radius
does not change significantly $\delta \log R_{\rm e}=0.07 \pm 0.07$.
Those trends are amplified if significant black hole growth is
assumed.

Although at this stage the measurement uncertainties do not allow firm
conclusions, it is clear that this empirical methodology holds great
promise as a way to disentangle evolution of the stellar populations,
black hole growth, and dynamical evolution of the host. In the future,
with larger samples covering a wider range of masses and redshifts, it
will be possible to study in the detail the co-evolution of spheroids
and black holes, perhaps in the framework or more fundamental
underlying correlations such as that proposed by
\citet{Hop++07}.

\section{Discussion and Conclusions}
\label{sec:disc}

In previous sections, we have presented evidence that just four
billion years ago, there was a population of supermassive black holes
living in smaller and less luminous spheroids than today.  Based on a
detailed analysis of uncertainties and selection effects, this
evolutionary signature appears significant at the 95\% CL. In this
section we will briefly discuss our results in the broader context of
galaxy formation and evolution and identify a possible physical
mechanism. However, before discussing the interpretation of this
result, it is worth mentioning two caveats (see also the discussion in
paper I). i) Our measurement relies on \mbh\, estimates based on an
empirically calibrated method. The sample of local calibrators is very
small and does not necessarily match exactly the properties of our
distant sample. More work on the local and distant samples remains to
be done before unknown systematics can be firmly ruled out. ii) The
local scaling relations are based on very limited samples.  The
quiescent sample consists of approximately 40 objects. In addition to
the intrinsic difficulty of resolving the sphere of influence, the
local sample is composed of earlier type galaxies than the sample
considered here. As discussed in paper I, the two samples may have an
evolutionary sequence built in at the selection.  In this respect, the
local AGN sample is very important, as the selection process and
properties are very similar to those of our own sample. However the
local AGN sample is comparable or smaller in size to our distant
sample, and therefore it is hard to identify subtle differences which
could point to some unknown selection effect.
 
Keeping these two caveats in the back of our minds, we move on to
discuss and interpret the main result of this paper, that black hole
mass growth appears to be completed before bulge growth. An important
clue is given by the morphology of the host galaxies, provided by the
HST images. The majority of galaxies in our sample are not
spheroid-dominated elliptical or lenticulars, but rather intermediate
or late type spirals. Most remarkably, as shown in Figure, 6/20
galaxies are observed to show signs of a major ongoing merging (3) or
to be morphologically disturbed (3). Considering only the three most
extreme cases and assuming that major mergers are visually
identifiable for $\sim$0.5-1 Gyr \citep{Cox++06a,Cox++06b}, this
finding implies that most or possibly all the galaxies in our sample
will undergo a major merger in between the time of observation and
today. Gas-rich mergers are believed to be the main mechanism
transforming stellar and gaseous disks into stellar spheroids, leading
to substantial increase in spheroid luminosity and velocity
dispersion. By converting rotation supported stars into pressure
supported stars, a merger with a disk dominated system (and hence
negligible supermassive black hole) could grow the bulge more
efficiently then the corresponding growth of the black hole by
accretion of cold gas \citep{Cro06}. 

As calculated in \S~\ref{sec:mani}, in order to satisfy the local
scaling relations, assuming that black hole growth is negligible, the
stellar mass of the spheroid will have to grow by approximately 60\%,
while the spheroid mass to light ratio wold have to remain
approximately constant, qualitatively consistent with a single
dissipative merger \citep{Cox++06a,Cox++06b}, and associated bursts of
star formation. This scenario is similar to that discussed in paper I,
to which the reader is referred for further discussion and references,
with two important additions: i) the high resolution HST images
provide direct evidence for gas rich mergers; ii) our joint analysis
of several scaling relations allows us to quantify the expected growth
of the spheroid mass and constrain the evolution of the stellar
populations of the spheroid via the evolution of the mass-to-light
ratio. Our imaging results may also suggest an apparent evolution of
Seyfert 1 host galaxy morphologies from $z=0.36$ to the current
epoch. This is because most local Seyfert 1 galaxies have substantial
bulges, with normal stellar populations, and no signs of interactions
\citep{Hun++99a}.  Although currently most Seyfert 1s at z=0 do not
show signs of ongoing mergers, they do have hints that these occurred
$\sim$ 1 Gyr earlier \citep{H+M99}, but that their host galaxy
morphologies have settled back down to normal by today
\citep{H+M04}. So our result appears to be consistent with the idea
that Seyfert host galaxies, as quiescent galaxies
\citep[e.g.][]{Bun++04,Lot++06a}, were more disturbed by interactions
4 Gyrs ago than they are today. Unfortunately our sample is too small
to make a proper comparison, and to derive statistically significant
results. However, we checked the consistency of this statement using
the GOODS database. The merger rate for our sample is 3/20
($0.13^{+0.11}_{-0.07}$) considering only close pairs/mergers, and it
increases to 6/20 considering all disturbed systems
($0.30\pm0.12$). As a control sample, we selected from GOODS all
galaxies with luminosity $17.6<i'<20.0$ (i.e. the same range in
stellar luminosity that our sample, excluding the point source), in
the redshift range $0.26<z<0.46$ (morphologies, photometric and
spectroscopic redshifts from Bundy et al.\ 2006) and performed the same
visual classification. We found 8/42 close pairs/mergers
($0.19\pm0.07$), increasing to 12/42 ($0.28\pm0.08$) if we consider
all disturbed systems. This is in good agreement with the fraction
observed for the distant Seyferts, and somewhat larger than observed
in the local universe \citep[e.g.][]{Pat++02}. The small size of our
sample does not warrant a more detailed attempt to measure the merger
fraction, so this is left for future work and larger samples.

Independent evidence appears to support the scenario discussed
above. From the galaxy evolution point of view, although the most
massive spheroids appear to have completed most of their growth by
$z\sim1$, it is clear that spheroids of a few $10^{10}$ M$_{\odot}$
are undergoing significant evolution at $z\sim0.4$. This view is
supported by the signatures of recent star formation detected by
Fundamental Plane studies \citep[e.g.,][and references
therein]{Tre++05b,vdW++05,dSA++05}, by the evolution of the mass
function (e.g. Bundy, Treu \& Ellis 2007 and references therein), and
by the evolution of the quenching or transition mass \citep[e.g.,][and
references therein]{Hop++07bundy}.  From the point of view of black
hole demographics, it is hard to pinpoint with sufficient precision
the growth of black holes in this mass range, although general
arguments based on the global star formation and accretion history
suggest that black hole growth may predate bulge growth
\citep[e.g.,][]{Mer04}. At larger masses, the existence of
impressively luminous quasars at $z\sim6$ \citep{Fan++06a} -- with the
high black hole to host galaxy mass ratio determined by radio
observations \citep{Wal++04a}, appears to be consistent with our
scenario, although it is hard to make a direct comparison, considering
that evolution may very well be mass-dependent.

Recent studies of the scaling relations between black hole mass and
host galaxy properties tend to focus on higher masses and redshifts
than are in our sample. For example, \citet{Pen++06qsob,Pen++06qsoa}
study the host galaxies of lensed quasars out to $z\sim4$, ruling out
pure luminosity evolution and finding that the ratio between \mbh\,
and $M_{\rm sph}$ was $\sim4$ times larger at $z\sim2-3$ than today.
Similar results are obtained by other studies of quasar host galaxies
at comparable redshifts \citep{Shi++06,McL++06a}. These results are
consistent with our own, although the comparison requires caution
considering that the higher redshift studies typically rely on UV
broad emission lines and fluxes for estimating \mbh, instead of
H$\beta$, and that the contrast between AGN and host light is more
unfavorable than in our case. This prevents accurate decomposition of
the host spheroid light and direct determination of the stellar
velocity dispersion, which needs to be inferred from proxies such as
CO and [\ion{O}{3}] line widths \citep{Bon++05,Sal++06,Wal++04a}. An
alternate approach followed by \citet{A+S05b} leads to the opposite
conclusion.  They use the correlation length of AGN hosts at
$z\sim2-3$ to estimate the virial mass of the halo, and the \ion{C}{4}
line width and UV flux at 1350\AA\, to estimate \mbh. They compare the
inferred relation between halo mass and black hole mass with the local
relation \citep{Fer02}, finding no evidence for evolution. Given their
error bars, they rule out evolution of over one order of magnitude in
the ratio -- i.e. evolution of the form $M_{\rm BH}/M_{\rm sph}\propto
(1+z)^{2.5}$, with $z\sim2.5$ -- at 90\% CL. However, they cannot rule
out evolution by a factor of 6 that would be predicted extrapolating
our best estimate.

In conclusion, it seems that several lines of evidence are beginning
to point in the same direction: black holes appear to complete their
growth before their host galaxies. The uncertainties are still large
and much work remains to be done before evolution can be considered
conclusively detected. However, with the advent of new technologies
such as laser guide star adaptive optics and robotic telescopes it
will hopefully be possible to improve dramatically our ability to
measure black hole masses in the local and distant universe, and thus
reduce the main source of error.

\section{Summary}
\label{sec:sum}

This paper is devoted to the study of the cosmic evolution of black
holes and their host galaxies. To this aim we have performed a detailed
analysis of a sample of 22 Seyfert 1 galaxies at $z\sim0.36$. The
choice of this particular redshift and moderate luminosity AGNs allows
us to derive precision measurements of the host galaxy properties, as
well as to obtain an estimate of the black hole mass from the dynamics
of the broad line region. Using high resolution images taken with ACS
we derived luminosity, effective radius and effective surface
brightness of the host spheroid as well as nuclear luminosity. We
combined this information with emission line widths and stellar
velocity dispersion based on high signal to noise Keck spectroscopy
(paper I) to construct the \mbh-L, \mbh-$\sigma$ and Fundamental Plane
relations of distant Seyferts. We compared the $z\sim0.36$ scaling
relations with those followed by local samples of quiescent and active
galaxies to determine evolutionary trends. The main results can be
summarized as follows:

\begin{enumerate}

\item The \mbh-L$_{\rm B}$ relation at $z\sim0.36$ is inconsistent
with the local relation and the assumption of pure luminosity
evolution of the host galaxy. Adopting pure luminosity evolution
consistent with Fundamental Plane studies, the offset from the local
relation corresponds to an offset in present day B-band luminosity of
$\Delta \log L_{\rm B,0}=0.40\pm0.11\pm0.15$, i.e. $\Delta \log M_{\rm
BH} = 0.51\pm0.14\pm0.19$, in the sense that black holes lived in
smaller bulges at $z\sim0.36$ than today.

\item The \mbh-$\sigma$ relation at $z\sim0.36$ is inconsistent with 
the local relation and the assumption of pure luminosity
evolution. The offset with respect to the local relation corresponds
to $\Delta \log \sigma = 0.13\pm0.03\pm0.05$, i.e. $\Delta \log M_{\rm
BH} = 0.54\pm0.12\pm0.21$, in the sense that black holes lived in
smaller bulges at $z\sim0.36$ than today.

\item Monte Carlo simulations show that the offset is not the 
result of selection effects, which are negligible for the
\mbh-$\sigma$ relation, and can account for at most 0.1 dex of the
observed offset of the \mbh-L relation.

\item In order to satisfy the local \mbh-$\sigma$, \mbh-L and FP
relations by $z=0$ -- assuming no black hole growth -- our distant
spheroids have to grow their stellar mass by approximately 60\%
($\Delta \log M_{\rm sp}=0.20\pm0.14$) in the next 4 billion years,
while preserving their size and holding their stellar mass to light
ratio approximately constant. This corresponds to an evolution of the
black hole to spheroid mass ratio of the form $M_{\rm BH} / M_{\rm
sph} \propto (1+z)^{1.5\pm1.0}$.

\end{enumerate}

Assuming that our results are not due to unknown systematic errors or
unknown selection effects, the observed evolution can be qualitatively
explained if our Seyferts undergo a single collisional merger with a
disk-dominated system between z=0.36 and today. This is consistent
with the observed merging/interacting fraction and a timescale for
merging visibility of $\sim$1 Gyr. A single merger could increase the
spheroid mass by transporting stellar mass from the progenitors disks,
without the corresponding growth of the central black holes due to the
lack of black hole in the disk dominated system.  At the same time,
this process would add younger stars to the spheroid (either from the
merging disks or from newly formed stars) thus counteracting the
fading of the old stellar populations and producing an approximately
constant stellar mass to light ratio in the spheroid. Numerical
simulations including realistic prescriptions for star formation, AGN
activity and mass loss will be needed to see if these mergers do,
indeed, preserve R$_{\rm e}$ and M$_{\rm sph}$/L$_{\rm B}$. If these
indications are supported by future studies, then they will confirm
that black holes completed their growth before their host galaxies and
are perhaps to be seen less as a by-product of galaxy formation than
as an orchestrator \citep{S+R98,Bla99}.

\acknowledgments
 
We thank David Hogg and the SDSS project for providing calibrated
images of the local comparison sample. We thank Aaron Barth, Brad
Peterson, Gregory Shields, and Risa Wechsler for discussions. We thank
Chien Peng for useful discussions and advice on using galfit and the
anonymous referee for useful suggestions and constructive
criticism. We are grateful to Kevin Bundy for providing the catalog of
galaxies in the GOODS fields with photometric and spectroscopic
redshift used to find the control sample described in
\S~\ref{sec:disc}. This work is based on data obtained with the Hubble
Space Telescope and the 10m W.M. Keck Telescope. We acknowledge
financial support from NASA through HST grant GO-10216. TT
acknowledges support from the NSF through CAREER award NSF-0642621,
and from the Sloan Foundation through a Sloan Research Fellowship.


\begin{thebibliography}{80}
\expandafter\ifx\csname natexlab\endcsname\relax\def\natexlab#1{#1}\fi

\bibitem[{{Adelberger} \& {Steidel}(2005)}]{A+S05b}
{Adelberger}, K.~L. \& {Steidel}, C.~C. 2005, \apjl, 627, L1

\bibitem[{{Bentz} {et~al.}(2006{\natexlab{a}}){Bentz}, {Denney}, {Cackett},
  {Dietrich}, {Fogel}, {Ghosh}, {Horne}, {Kuehn}, {Minezaki}, {Onken},
  {Peterson}, {Pogge}, {Pronik}, {Richstone}, {Sergeev}, {Vestergaard},
  {Walker}, \& {Yoshii}}]{Ben++06b}
{Bentz}, M.~C., {Denney}, K.~D., {Cackett}, E.~M., {Dietrich}, M., {Fogel},
  J.~K.~J., {Ghosh}, H., {Horne}, K., {Kuehn}, C., {Minezaki}, T., {Onken},
  C.~A., {Peterson}, B.~M., {Pogge}, R.~W., {Pronik}, V.~I., {Richstone},
  D.~O., {Sergeev}, S.~G., {Vestergaard}, M., {Walker}, M.~G., \& {Yoshii}, Y.
  2006{\natexlab{a}}, \apj, 651, 775

\bibitem[{{Bentz} {et~al.}(2006{\natexlab{b}}){Bentz}, {Peterson}, {Pogge},
  {Vestergaard}, \& {Onken}}]{Ben++06a}
{Bentz}, M.~C., {Peterson}, B.~M., {Pogge}, R.~W., {Vestergaard}, M., \&
  {Onken}, C.~A. 2006{\natexlab{b}}, \apj, 644, 133

\bibitem[{{Bernardi} {et~al.}(2006){Bernardi}, {Sheth}, {Tundo}, \&
  {Hyde}}]{Ber++06b}
{Bernardi}, M., {Sheth}, R.~K., {Tundo}, E., \& {Hyde}, J.~B. 2006, ArXiv
  Astrophysics e-prints

\bibitem[{{Blandford}(1999)}]{Bla99}
{Blandford}, R.~D. 1999, 87

\bibitem[{{Bonning} {et~al.}(2005){Bonning}, {Shields}, {Salviander}, \&
  {McLure}}]{Bon++05}
{Bonning}, E.~W., {Shields}, G.~A., {Salviander}, S., \& {McLure}, R.~J. 2005,
  \apj, 626, 89

\bibitem[{{Boylan-Kolchin} {et~al.}(2006){Boylan-Kolchin}, {Ma}, \&
  {Quataert}}]{BMQ06}
{Boylan-Kolchin}, M., {Ma}, C.-P., \& {Quataert}, E. 2006, \mnras, 369, 1081

\bibitem[{{Bundy} {et~al.}(2004){Bundy}, {Fukugita}, {Ellis}, {Kodama}, \&
  {Conselice}}]{Bun++04}
{Bundy}, K., {Fukugita}, M., {Ellis}, R.~S., {Kodama}, T., \& {Conselice},
  C.~J. 2004, \apjl, 601, L123

\bibitem[Bundy et al.(2006)]{2006ApJ...651..120B} Bundy, K., et al.\ 2006, \apj, 651, 120 

\bibitem[Bundy et al.(2007)]{2007arXiv0705.1007B} Bundy, K., Treu, T., \& Ellis, R.~S.\ 2007, ArXiv e-prints, 705, arXiv:0705.1007 

\bibitem[{{Ciotti} \& {Ostriker}(2007)}]{C+O07}
{Ciotti}, L. \& {Ostriker}, J.~P. 2007, ArXiv e-prints, arXiv:astro-ph/0703057

\bibitem[{{Ciotti} \& {van Albada}(2001)}]{C+V01}
{Ciotti}, L. \& {van Albada}, T.~S. 2001, \apjl, 552, L13

\bibitem[Collin et al.\ (2006)]{Col++06}{Collin}, S., {Kawaguchi}, T., {Peterson}, B.~M. \& {Vestergaard}, M. 2006, \aap, 456, 75

\bibitem[{{Cox} {et~al.}(2006{\natexlab{a}}){Cox}, {Dutta}, {Di Matteo},
  {Hernquist}, {Hopkins}, {Robertson}, \& {Springel}}]{Cox++06b}
{Cox}, T.~J., {Dutta}, S.~N., {Di Matteo}, T., {Hernquist}, L., {Hopkins},
  P.~F., {Robertson}, B., \& {Springel}, V. 2006{\natexlab{a}}, \apj, 650, 791

\bibitem[{{Cox} {et~al.}(2006{\natexlab{b}}){Cox}, {Jonsson}, {Primack}, \&
  {Somerville}}]{Cox++06a}
{Cox}, T.~J., {Jonsson}, P., {Primack}, J.~R., \& {Somerville}, R.~S.
  2006{\natexlab{b}}, \mnras, 373, 1013

\bibitem[{{Croton}(2006)}]{Cro06}
{Croton}, D.~J. 2006, \mnras, 369, 1808

\bibitem[{{Croton} {et~al.}(2006){Croton}, {Springel}, {White}, {De Lucia},
  {Frenk}, {Gao}, {Jenkins}, {Kauffmann}, {Navarro}, \& {Yoshida}}]{Cro++06b}
{Croton}, D.~J., {Springel}, V., {White}, S.~D.~M., {De Lucia}, G., {Frenk},
  C.~S., {Gao}, L., {Jenkins}, A., {Kauffmann}, G., {Navarro}, J.~F., \&
  {Yoshida}, N. 2006, \mnras, 365, 11

\bibitem[{{De Lucia} \& {Blaizot}(2007)}]{D+B07}
{De Lucia}, G. \& {Blaizot}, J. 2007, \mnras, 375, 2

\bibitem[{{de Vaucouleurs}(1948)}]{deV48}
{de Vaucouleurs}, G. 1948, Annales d'Astrophysique, 11, 247

\bibitem[{{Di Matteo} {et~al.}(2005){Di Matteo}, {Springel}, \&
  {Hernquist}}]{DSH05}
{Di Matteo}, T., {Springel}, V., \& {Hernquist}, L. 2005, \nat, 433, 604

\bibitem[{{di Serego Alighieri} {et~al.}(2005){di Serego Alighieri}, {Vernet},
  {Cimatti}, {Lanzoni}, {Cassata}, {Ciotti}, {Daddi}, {Mignoli}, {Pignatelli},
  {Pozzetti}, {Renzini}, {Rettura}, \& {Zamorani}}]{dSA++05}
{di Serego Alighieri}, S., {Vernet}, J., {Cimatti}, A., {Lanzoni}, B.,
  {Cassata}, P., {Ciotti}, L., {Daddi}, E., {Mignoli}, M., {Pignatelli}, E.,
  {Pozzetti}, L., {Renzini}, A., {Rettura}, A., \& {Zamorani}, G. 2005, \aap,
  442, 125

\bibitem[{{Driver} {et~al.}(2007){Driver}, {Allen}, {Liske}, \&
  {Graham}}]{Dri++07}
{Driver}, S.~P., {Allen}, P.~D., {Liske}, J., \& {Graham}, A.~W. 2007, \apjl,
  657, L85

\bibitem[{{Fan} {et~al.}(2006){Fan}, {Strauss}, {Richards}, {Hennawi},
  {Becker}, {White}, {Diamond-Stanic}, {Donley}, {Jiang}, {Kim}, {Vestergaard},
  {Young}, {Gunn}, {Lupton}, {Knapp}, {Schneider}, {Brandt}, {Bahcall},
  {Barentine}, {Brinkmann}, {Brewington}, {Fukugita}, {Harvanek}, {Kleinman},
  {Krzesinski}, {Long}, {Neilsen}, {Nitta}, {Snedden}, \& {Voges}}]{Fan++06a}
{Fan}, X., {Strauss}, M.~A., {Richards}, G.~T., {Hennawi}, J.~F., {Becker},
  R.~H., {White}, R.~L., {Diamond-Stanic}, A.~M., {Donley}, J.~L., {Jiang}, L.,
  {Kim}, J.~S., {Vestergaard}, M., {Young}, J.~E., {Gunn}, J.~E., {Lupton},
  R.~H., {Knapp}, G.~R., {Schneider}, D.~P., {Brandt}, W.~N., {Bahcall}, N.~A.,
  {Barentine}, J.~C., {Brinkmann}, J., {Brewington}, H.~J., {Fukugita}, M.,
  {Harvanek}, M., {Kleinman}, S.~J., {Krzesinski}, J., {Long}, D., {Neilsen},
  Jr., E.~H., {Nitta}, A., {Snedden}, S.~A., \& {Voges}, W. 2006, \aj, 131,
  1203

\bibitem[{{Ferrarese}(2002)}]{Fer02}
{Ferrarese}, L. 2002, \apj, 578, 90

\bibitem[{{Ferrarese} \& {Ford}(2005)}]{F+F05}
{Ferrarese}, L. \& {Ford}, H. 2005, Space Science Reviews, 116, 523

\bibitem[{{Ferrarese} \& {Merritt}(2000)}]{F+M00}
{Ferrarese}, L. \& {Merritt}, D. 2000, \apjl, 539, L9

\bibitem[{{Gebhardt} {et~al.}(2000){Gebhardt}, {Bender}, {Bower}, {Dressler},
  {Faber}, {Filippenko}, {Green}, {Grillmair}, {Ho}, {Kormendy}, {Lauer},
  {Magorrian}, {Pinkney}, {Richstone}, \& {Tremaine}}]{Geb++00}
{Gebhardt}, K., {Bender}, R., {Bower}, G., {Dressler}, A., {Faber}, S.~M.,
  {Filippenko}, A.~V., {Green}, R., {Grillmair}, C., {Ho}, L.~C., {Kormendy},
  J., {Lauer}, T.~R., {Magorrian}, J., {Pinkney}, J., {Richstone}, D., \&
  {Tremaine}, S. 2000, \apjl, 539, L13

\bibitem[{{Gebhardt} {et~al.}(2001){Gebhardt}, {Lauer}, {Kormendy}, {Pinkney},
  {Bower}, {Green}, {Gull}, {Hutchings}, {Kaiser}, {Nelson}, {Richstone}, \&
  {Weistrop}}]{Geb++01}
{Gebhardt}, K., {Lauer}, T.~R., {Kormendy}, J., {Pinkney}, J., {Bower}, G.~A.,
  {Green}, R., {Gull}, T., {Hutchings}, J.~B., {Kaiser}, M.~E., {Nelson},
  C.~H., {Richstone}, D., \& {Weistrop}, D. 2001, \aj, 122, 2469

\bibitem[{{Graham} \& {Driver}(2007)}]{G+D07}
{Graham}, A.~W. \& {Driver}, S.~P. 2007, \apj, 655, 77

\bibitem[{{Granato} {et~al.}(2004){Granato}, {De Zotti}, {Silva}, {Bressan}, \&
  {Danese}}]{Gra++04}
{Granato}, G.~L., {De Zotti}, G., {Silva}, L., {Bressan}, A., \& {Danese}, L.
  2004, \apj, 600, 580

\bibitem[{{Greene} \& {Ho}(2006)}]{G+H06ms}
{Greene}, J.~E. \& {Ho}, L.~C. 2006, \apjl, 641, L21

\bibitem[{{Greene} \& {Ho}(2007)}]{G+H07}
{Greene}, J.~E. \& {Ho}, L.~C. 2007, ArXiv Astrophysics e-prints, 705,
arXiv:0705.0020

\bibitem[{{Haehnelt} \& {Kauffmann}(2000)}]{H+K00}
{Haehnelt}, M.~G. \& {Kauffmann}, G. 2000, \mnras, 318, L35

\bibitem[{{H{\"a}ring} \& {Rix}(2004)}]{H+R04}
{H{\"a}ring}, N. \& {Rix}, H.-W. 2004, \apjl, 604, L89

\bibitem[Hopkins et al.(2007)]{Hop++07bundy} Hopkins, P.~F., Bundy, K., Hernquist, L., \& Ellis, R.~S.\ 2007, \apj, 659, 976 

\bibitem[{{Hopkins} {et~al.}(2006{\natexlab{b}}){Hopkins}, {Hernquist}, {Cox},
  {Di Matteo}, {Robertson}, \& {Springel}}]{Hop++06c}
{Hopkins}, P.~F., {Hernquist}, L., {Cox}, T.~J., {Di Matteo}, T., {Robertson},
  B., \& {Springel}, V. 2006{\natexlab{b}}, \apjs, 163, 1

\bibitem[{{Hopkins} {et~al.}(2007){Hopkins}, {Hernquist}, {Cox}, {Robertson},
  \& {Krause}}]{Hop++07}
{Hopkins}, P.~F., {Hernquist}, L., {Cox}, T.~J., {Robertson}, B., \& {Krause},
  E. 2007, ArXiv Astrophysics e-prints

\bibitem[{{Hunt} \& {Malkan}(1999)}]{H+M99}
{Hunt}, L.~K. \& {Malkan}, M.~A. 1999, \apj, 516, 660

\bibitem[{{Hunt} \& {Malkan}(2004)}]{H+M04}
---. 2004, \apj, 616, 707

\bibitem[{{Hunt} {et~al.}(1999){Hunt}, {Malkan}, {Moriondo}, \&
  {Salvati}}]{Hun++99a}
{Hunt}, L.~K., {Malkan}, M.~A., {Moriondo}, G., \& {Salvati}, M. 1999, \apj,
  510, 637

\bibitem[{{Kaspi} {et~al.}(2005){Kaspi}, {Maoz}, {Netzer}, {Peterson},
  {Vestergaard}, \& {Jannuzi}}]{Kas++05}
{Kaspi}, S., {Maoz}, D., {Netzer}, H., {Peterson}, B.~M., {Vestergaard}, M., \&
  {Jannuzi}, B.~T. 2005, \apj, 629, 61

\bibitem[{{Kaspi} {et~al.}(2000){Kaspi}, {Smith}, {Netzer}, {Maoz}, {Jannuzi},
  \& {Giveon}}]{Kas++00a}
{Kaspi}, S., {Smith}, P.~S., {Netzer}, H., {Maoz}, D., {Jannuzi}, B.~T., \&
  {Giveon}, U. 2000, \apj, 533, 631

\bibitem[{{Kormendy} \& {Richstone}(1995)}]{K+R95}
{Kormendy}, J. \& {Richstone}, D. 1995, \araa, 33, 581

\bibitem[{{Lotz} {et~al.}(2006){Lotz}, {Davis}, {Faber}, {Guhathakurta},
  {Gwyn}, {Huang}, {Koo}, {Le Floc'h}, {Lin}, {Newman}, {Noeske}, {Papovich},
  {Willmer}, {Coil}, {Conselice}, {Cooper}, {Hopkins}, {Metevier}, {Primack},
  {Rieke}, \& {Weiner}}]{Lot++06a}
{Lotz}, J.~M., {Davis}, M., {Faber}, S.~M., {Guhathakurta}, P., {Gwyn}, S.,
  {Huang}, J., {Koo}, D.~C., {Le Floc'h}, E., {Lin}, L., {Newman}, J.,
  {Noeske}, K., {Papovich}, C., {Willmer}, C.~N.~A., {Coil}, A., {Conselice},
  C.~J., {Cooper}, M., {Hopkins}, A.~M., {Metevier}, A., {Primack}, J.,
  {Rieke}, G., \& {Weiner}, B.~J. 2006, ArXiv Astrophysics e-prints

\bibitem[{{Magorrian} {et~al.}(1998){Magorrian}, {Tremaine}, {Richstone},
  {Bender}, {Bower}, {Dressler}, {Faber}, {Gebhardt}, {Green}, {Grillmair},
  {Kormendy}, \& {Lauer}}]{Mag++98}
{Magorrian}, J., {Tremaine}, S., {Richstone}, D., {Bender}, R., {Bower}, G.,
  {Dressler}, A., {Faber}, S.~M., {Gebhardt}, K., {Green}, R., {Grillmair}, C.,
  {Kormendy}, J., \& {Lauer}, T. 1998, \aj, 115, 2285

\bibitem[{{Malbon} {et~al.}(2006){Malbon}, {Baugh}, {Frenk}, \&
  {Lacey}}]{Mal++06}
{Malbon}, R.~K., {Baugh}, C.~M., {Frenk}, C.~S., \& {Lacey}, C.~G. 2006, ArXiv
  Astrophysics e-prints, astro-ph/0607424 

\bibitem[{{Marconi} \& {Hunt}(2003)}]{M+H03}
{Marconi}, A. \& {Hunt}, L.~K. 2003, \apjl, 589, L21

\bibitem[{{McLure} {et~al.}(2006){McLure}, {Jarvis}, {Targett}, {Dunlop}, \&
  {Best}}]{McL++06a}
{McLure}, R.~J., {Jarvis}, M.~J., {Targett}, T.~A., {Dunlop}, J.~S., \& {Best},
  P.~N. 2006, \mnras, 368, 1395

\bibitem[{{Merloni}(2004)}]{Mer04}
{Merloni}, A. 2004, \mnras, 353, 1035

\bibitem[{{Miralda-Escud{\'e}} \& {Kollmeier}(2005)}]{M+K05}
{Miralda-Escud{\'e}}, J. \& {Kollmeier}, J.~A. 2005, \apj, 619, 30

\bibitem[{{Netzer} \& {Trakhtenbrot}(2007)}]{N+T07}{Netzer}, H. \& {Trakhtenbrot}, B. 2007, \apj, 654, 754

\bibitem[{{Nipoti} {et~al.}(2003){Nipoti}, {Londrillo}, \& {Ciotti}}]{NLC03}
{Nipoti}, C., {Londrillo}, P., \& {Ciotti}, L. 2003, \mnras, 342, 501

\bibitem[{{Novak} {et~al.}(2006){Novak}, {Faber}, \& {Dekel}}]{NFD06}
{Novak}, G.~S., {Faber}, S.~M., \& {Dekel}, A. 2006, \apj, 637, 96

\bibitem[{{Onken} {et~al.}(2004){Onken}, {Ferrarese}, {Merritt}, {Peterson},
  {Pogge}, {Vestergaard}, \& {Wandel}}]{Onk++04}
{Onken}, C.~A., {Ferrarese}, L., {Merritt}, D., {Peterson}, B.~M., {Pogge},
  R.~W., {Vestergaard}, M., \& {Wandel}, A. 2004, \apj, 615, 645

\bibitem[{{Patton} {et~al.}(2002){Patton}, {Pritchet}, {Carlberg}, {Marzke},
  {Yee}, {Hall}, {Lin}, {Morris}, {Sawicki}, {Shepherd}, \& {Wirth}}]{Pat++02}
{Patton}, D.~R., {Pritchet}, C.~J., {Carlberg}, R.~G., {Marzke}, R.~O., {Yee},
  H.~K.~C., {Hall}, P.~B., {Lin}, H., {Morris}, S.~L., {Sawicki}, M.,
  {Shepherd}, C.~W., \& {Wirth}, G.~D. 2002, \apj, 565, 208

\bibitem[{{Peng} {et~al.}(2002){Peng}, {Ho}, {Impey}, \& {Rix}}]{Pen++02}
{Peng}, C.~Y., {Ho}, L.~C., {Impey}, C.~D., \& {Rix}, H.-W. 2002, \aj, 124, 266

\bibitem[{{Peng} {et~al.}(2006{\natexlab{a}}){Peng}, {Impey}, {Ho}, {Barton},
  \& {Rix}}]{Pen++06qsoa}
{Peng}, C.~Y., {Impey}, C.~D., {Ho}, L.~C., {Barton}, E.~J., \& {Rix}, H.-W.
  2006{\natexlab{a}}, \apj, 640, 114

\bibitem[{{Peng} {et~al.}(2006{\natexlab{b}}){Peng}, {Impey}, {Rix},
  {Kochanek}, {Keeton}, {Falco}, {Leh{\'a}r}, \& {McLeod}}]{Pen++06qsob}
{Peng}, C.~Y., {Impey}, C.~D., {Rix}, H.-W., {Kochanek}, C.~S., {Keeton},
  C.~R., {Falco}, E.~E., {Leh{\'a}r}, J., \& {McLeod}, B.~A.
  2006{\natexlab{b}}, \apj, 649, 616

\bibitem[{{Peterson}(2007)}]{Pet07}
{Peterson}, B.~M. 2007, ArXiv Astrophysics e-prints

\bibitem[{{Peterson} {et~al.}(2004){Peterson}, {Ferrarese}, {Gilbert}, {Kaspi},
  {Malkan}, {Maoz}, {Merritt}, {Netzer}, {Onken}, {Pogge}, {Vestergaard}, \&
  {Wandel}}]{Pet++04}
{Peterson}, B.~M., {Ferrarese}, L., {Gilbert}, K.~M., {Kaspi}, S., {Malkan},
  M.~A., {Maoz}, D., {Merritt}, D., {Netzer}, H., {Onken}, C.~A., {Pogge},
  R.~W., {Vestergaard}, M., \& {Wandel}, A. 2004, \apj, 613, 682

\bibitem[{{Robertson} {et~al.}(2006){Robertson}, {Hernquist}, {Cox}, {Di
  Matteo}, {Hopkins}, {Martini}, \& {Springel}}]{Rob++06a}
{Robertson}, B., {Hernquist}, L., {Cox}, T.~J., {Di Matteo}, T., {Hopkins},
  P.~F., {Martini}, P., \& {Springel}, V. 2006, \apj, 641, 90

\bibitem[{{Salviander} {et~al.}(2006){Salviander}, {Shields}, {Gebhardt}, \&
  {Bonning}}]{Sal++06}
{Salviander}, S., {Shields}, G.~A., {Gebhardt}, K., \& {Bonning}, E.~W. 2006,
  ApJ, in press, astro-ph/0612568

\bibitem[{{Sersic}(1968)}]{Ser68}
{Sersic}, J.~L. 1968

\bibitem[{{Sheth} {et~al.}(2003){Sheth}, {Bernardi}, {Schechter}, {Burles},
  {Eisenstein}, {Finkbeiner}, {Frieman}, {Lupton}, {Schlegel}, {Subbarao},
  {Shimasaku}, {Bahcall}, {Brinkmann}, \& {Ivezi{\'c}}}]{Shet++03}
{Sheth}, R.~K., {Bernardi}, M., {Schechter}, P.~L., {Burles}, S., {Eisenstein},
  D.~J., {Finkbeiner}, D.~P., {Frieman}, J., {Lupton}, R.~H., {Schlegel},
  D.~J., {Subbarao}, M., {Shimasaku}, K., {Bahcall}, N.~A., {Brinkmann}, J., \&
  {Ivezi{\'c}}, {\v Z}. 2003, \apj, 594, 225

\bibitem[{{Shields} {et~al.}(2003){Shields}, {Gebhardt}, {Salviander}, {Wills},
  {Xie}, {Brotherton}, {Yuan}, \& {Dietrich}}]{Shi++03}
{Shields}, G.~A., {Gebhardt}, K., {Salviander}, S., {Wills}, B.~J., {Xie}, B.,
  {Brotherton}, M.~S., {Yuan}, J., \& {Dietrich}, M. 2003, \apj, 583, 124

\bibitem[{{Shields} {et~al.}(2006){Shields}, {Menezes}, {Massart}, \& {Vanden
  Bout}}]{Shi++06}
{Shields}, G.~A., {Menezes}, K.~L., {Massart}, C.~A., \& {Vanden Bout}, P.
  2006, \apj, 641, 683

\bibitem[{{Silk} \& {Rees}(1998)}]{S+R98}
{Silk}, J. \& {Rees}, M.~J. 1998, \aap, 331, L1

\bibitem[{{Tremaine} {et~al.}(2002){Tremaine}, {Gebhardt}, {Bender}, {Bower},
  {Dressler}, {Faber}, {Filippenko}, {Green}, {Grillmair}, {Ho}, {Kormendy},
  {Lauer}, {Magorrian}, {Pinkney}, \& {Richstone}}]{Trem++02}
{Tremaine}, S., {Gebhardt}, K., {Bender}, R., {Bower}, G., {Dressler}, A.,
  {Faber}, S.~M., {Filippenko}, A.~V., {Green}, R., {Grillmair}, C., {Ho},
  L.~C., {Kormendy}, J., {Lauer}, T.~R., {Magorrian}, J., {Pinkney}, J., \&
  {Richstone}, D. 2002, \apj, 574, 740

\bibitem[{{Treu} {et~al.}(2005{\natexlab{a}}){Treu}, {Ellis}, {Liao}, \& {van
  Dokkum}}]{Tre++05a}
{Treu}, T., {Ellis}, R.~S., {Liao}, T.~X., \& {van Dokkum}, P.~G.
  2005{\natexlab{a}}, \apjl, 622, L5

\bibitem[{{Treu} {et~al.}(2005{\natexlab{b}}){Treu}, {Ellis}, {Liao}, {van
  Dokkum}, {Tozzi}, {Coil}, {Newman}, {Cooper}, \& {Davis}}]{Tre++05b}
{Treu}, T., {Ellis}, R.~S., {Liao}, T.~X., {van Dokkum}, P.~G., {Tozzi}, P.,
  {Coil}, A., {Newman}, J., {Cooper}, M.~C., \& {Davis}, M. 2005{\natexlab{b}},
  \apj, 633, 174

\bibitem[{{Treu} {et~al.}(2004){Treu}, {Malkan}, \& {Blandford}}]{TMB04}
{Treu}, T., {Malkan}, M.~A., \& {Blandford}, R.~D. 2004, \apjl, 615, L97

\bibitem[{{Treu} {et~al.}(2001{\natexlab{a}}){Treu}, {Stiavelli}, {Bertin},
  {Casertano}, \& {M{\o}ller}}]{Tre++01b}
{Treu}, T., {Stiavelli}, M., {Bertin}, G., {Casertano}, S., \& {M{\o}ller}, P.
  2001{\natexlab{a}}, \mnras, 326, 237

\bibitem[{{Treu} {et~al.}(2001{\natexlab{b}}){Treu}, {Stiavelli}, {M{\o}ller},
  {Casertano}, \& {Bertin}}]{Tre++01a}
{Treu}, T., {Stiavelli}, M., {M{\o}ller}, P., {Casertano}, S., \& {Bertin}, G.
  2001{\natexlab{b}}, \mnras, 326, 221

\bibitem[{{van der Wel} {et~al.}(2005){van der Wel}, {Franx}, {van Dokkum},
  {Rix}, {Illingworth}, \& {Rosati}}]{vdW++05}
{van der Wel}, A., {Franx}, M., {van Dokkum}, P.~G., {Rix}, H.-W.,
  {Illingworth}, G.~D., \& {Rosati}, P. 2005, \apj, 631, 145

\bibitem[{{Vestergaard}(2002)}]{Ves02}
{Vestergaard}, M. 2002, \apj, 571, 733

\bibitem[{{Vestergaard} \& {Peterson}(2006)}]{V+P06}
{Vestergaard}, M. \& {Peterson}, B.~M. 2006, \apj, 641, 689

\bibitem[{{Walter} {et~al.}(2004){Walter}, {Carilli}, {Bertoldi}, {Menten},
  {Cox}, {Lo}, {Fan}, \& {Strauss}}]{Wal++04a}
{Walter}, F., {Carilli}, C., {Bertoldi}, F., {Menten}, K., {Cox}, P., {Lo},
  K.~Y., {Fan}, X., \& {Strauss}, M.~A. 2004, \apjl, 615, L17

\bibitem[{{Wandel} {et~al.}(1999){Wandel}, {Peterson}, \& {Malkan}}]{WPM99}
{Wandel}, A., {Peterson}, B.~M., \& {Malkan}, M.~A. 1999, \apj, 526, 579

\bibitem[{{Woo} {et~al.}(2007){Woo}, {Treu}, {Malkan}, {Ferry}, \&
  {Misch}}]{Woo++07}
{Woo}, J.-H., {Treu}, T., {Malkan}, M.~A., {Ferry}, M.~A., \& {Misch}, T. 2007, ApJ, 661, 60

\bibitem[{{Woo} {et~al.}(2006){Woo}, {Treu}, {Malkan}, \&
  {Blandford}}]{Woo++06}
{Woo}, J.-H., {Treu}, T., {Malkan}, M.~A., \& {Blandford}, R.~D. 2006, \apj,
  645, 900

\bibitem[{{Woo} {et~al.}(2004){Woo}, {Urry}, {Lira}, {van der Marel}, \&
  {Maza}}]{Woo++04}
{Woo}, J.-H., {Urry}, C.~M., {Lira}, P., {van der Marel}, R.~P., \& {Maza}, J.
  2004, \apj, 617, 903

\bibitem[{{Woo} {et~al.}(2005){Woo}, {Urry}, {van der Marel}, {Lira}, \&
  {Maza}}]{Woo++05}
{Woo}, J.-H., {Urry}, C.~M., {van der Marel}, R.~P., {Lira}, P., \& {Maza}, J.
  2005, \apj, 631, 762

\end{thebibliography}

%

\clearpage

\begin{deluxetable}{lrrcrrr}
\tablewidth{0pt}
\tablecaption{Sample properties}
\tablehead{
\colhead{Name}        &
\colhead{RA (J2000)}     &
\colhead{DEC (J2000)}    &
\colhead{z}           &
\colhead{i'}    &
\colhead{$\sigma$}    
\\
\colhead{(1)} &
\colhead{(2)} &
\colhead{(3)} &
\colhead{(4)} &
\colhead{(5)} &
\colhead{(6)} &}
\tablecolumns{6}
\startdata
S01 &  15 39 16.23 & +03 23 22.06 & 0.3596       &  18.74 & $132\pm8$ \\
S02 &  16 11 11.67 & +51 31 31.12 & 0.3544\tablenotemark{a} &  18.94 & - \\
S03 &  17 32 03.11 & +61 17 51.96 & 0.3583\tablenotemark{a} &  18.20 & - \\
S04 &  21 02 11.51 & -06 46 45.03 & 0.3580       &  18.41 & 186$\pm$ 8 \\
S05 &  21 04 51.85 & -07 12 09.45 & 0.3531       &  18.35 & 132$\pm$ 5 \\
S06 &  21 20 34.19 & -06 41 22.24 & 0.3689       &  18.41 & 169$\pm$14 \\
S07 &  23 09 46.14 & +00 00 48.91 & 0.3520       &  18.11 & 145$\pm$13 \\
S08 &  23 59 53.44 & -09 36 55.53 & 0.3591       &  18.42 & 187$\pm$11 \\
S09 &  00 59 16.11 & +15 38 16.08 & 0.3548       &  18.16 & 187$\pm$15 \\
S10 &  01 01 12.07 & -09 45 00.76 & 0.3506\tablenotemark{a} &  17.92 & -          \\
S11 &  01 07 15.97 & -08 34 29.40 & 0.3562       &  18.34 & 127$\pm$ 9 \\
S12 &  02 13 40.60 & +13 47 56.06 & 0.3575       &  18.12 & 173$\pm$22 \\
S16 &  11 19 37.58 & +00 56 20.42 & 0.3702\tablenotemark{a} &  19.22 & -          \\
S21 &  11 05 56.18 & +03 12 43.26 & 0.3534\tablenotemark{a} &  17.21 & -          \\
S23 &  14 00 16.66 & -01 08 22.19 & 0.3515       &  18.08 & 172$\pm$ 8 \\
S24 &  14 00 34.71 & +00 47 33.48 & 0.3621       &  18.21 & 214$\pm$10 \\
S26 &  15 29 22.26 & +59 28 54.56 & 0.3691       &  18.88 & 128$\pm$ 8 \\
S27 &  15 36 51.28 & +54 14 42.71 & 0.3667\tablenotemark{a} &  18.80 & -          \cr
S28 &  16 11 56.30 & +45 16 11.04 & 0.3682       &  18.59 & 210$\pm$10 \cr
S29 &  21 58 41.93 & -01 15 00.33 & 0.3575\tablenotemark{a} &  18.77 & - \cr
S31 &  10 15 27.26 & +62 59 11.51 & 0.3504\tablenotemark{a} &  18.14 & -          \cr
S99 &  16 00 02.80 & +41 30 27.00 & 0.3690       &  18.33 & 224$\pm$12 \cr
\enddata
\label{tab:sample}
\tablecomments{
Col. (1): Target ID.  
Col. (2): RA.
Col. (3): DEC.
Col. (4): Redshift from stellar absorption lines.
Col. (5): Extinction corrected $i'$ AB magnitude from SDSS photometry.
Col. (6): Stellar velocity dispersion in \kms\, from paper I.}
\tablenotetext{a}{redshift from SDSS DR4.}
\end{deluxetable}

\begin{deluxetable}{lrrcrrrrc}
\tablewidth{0pt}
\tablecaption{New measurements}
\tablehead{
\colhead{Name}        &
\colhead{ i'  (total)}     &
\colhead{ i'  (spheroid)}    &
\colhead{$\log L_{\rm B}/L_{\rm \odot, B}$}           &
\colhead{SB$_{\rm e,B}$}    &
\colhead{R$_{\rm e}$}    &
\colhead{L$_{\rm 5100}$}  &  
\colhead{f$_{\rm nuc}$}  &
\colhead{$\log M_{\rm BH} / M_{\rm \odot}$} 
\\
\colhead{(1)} &
\colhead{(2)} &
\colhead{(3)} &
\colhead{(4)} &
\colhead{(5)} &
\colhead{(6)} &
\colhead{(7)} &
\colhead{(8)} &
\colhead{(9)}}
\tablecolumns{9}
\startdata
S01     & 18.50 & 19.92 & 10.28 & 21.85 & 5.29 & 0.74 & 0.29 & 8.21 \\
S02     & 19.03 & 19.85 & 10.31 & 20.27 & 2.63 & 0.36 & 0.22 & 7.99 \\
S03\tablenotemark{a} & 17.94 & 20.23 & 10.16 & 17.04 & 0.50 & 1.69 & 0.39 & 8.29 \\
S04     & 18.06 & 20.12 & 10.20 & 18.36 & 0.96 & 1.42 & 0.36 & 8.45 \\
S05     & 17.93 & 20.45 & 10.07 & 18.84 & 1.03 & 2.04 & 0.47 & 8.77 \\
S06     & 18.35 & 20.48 & 10.06 & 18.81 & 1.01 & 0.54 & 0.18 & 8.17 \\
S07     & 17.79 & 20.35 & 10.11 & 18.69 & 1.01 & 2.26 & 0.45 & 8.55 \\
S08     & 18.31 & 21.75 &  9.55 & 20.50 & 1.23 & 1.25 & 0.40 & 8.10 \\
S09     & 18.17 & 19.00 & 10.65 & 19.87 & 3.24 & 0.78 & 0.22 & 8.15 \\
S10\tablenotemark{a} & 18.01 & 19.30 & 10.53 & 16.08 & 0.49 & 1.11 & 0.27 & 8.27 \\
S12\tablenotemark{a}  & 18.12 & 21.16 &  9.78 & 18.14 & 0.54 & 1.05 & 0.28 & 8.69 \\
S16     & 19.14 & 22.26 &  9.34 & 19.97 & 0.76 & 0.70 & 0.48 & 8.26 \\
S21\tablenotemark{a} & 17.45 & 18.95 & 10.67 & 15.82 & 0.51 & 2.30 & 0.34 & 8.81 \cr
S23\tablenotemark{a} & 17.99 & 20.85 &  9.91 & 17.93 & 0.57 & 1.20 & 0.29 & 8.72 \cr
S24     & 18.06 & 18.59 & 10.81 & 22.41 & 12.6 & 0.44 & 0.11 & 8.33 \cr
S26     & 18.87 & 20.06 & 10.23 & 17.73 & 0.75 & 0.50 & 0.27 & 8.01 \cr
S27     & 18.51 & 19.46 & 10.46 & 21.17 & 4.78 & 0.92 & 0.36 & 8.10 \cr
\enddata
\label{tab:meas}
\tablecomments{
Col. (1): Target ID.  
Col. (2): Total extinction corrected F775W AB magnitude. 
Col. (3): Spheroid extinction corrected F775W AB magnitude.
Col. (4): Log$_{10}$ of spheroid luminosity in rest frame B (solar units), not corrected for evolution. Errors are estimated to be 0.2 dex.
Col. (5): Spheroid effective surface brightness in rest frame B (AB magnitudes arcsec$^{-2}$; see~\S~\ref{sssec:surf} for error discussion). 
Col. (6): Spheroid effective radius (kpc; see~\S~\ref{sssec:surf} for error discussion).
Col. (7): Nuclear luminosity at 5100 \AA\, (10$^{44}$ erg s$^{-1}$). Errors are estimated to be 20\%.
Col. (8): Nuclear light fraction in F775W. Errors are estimated to be 20\%.
Col. (9): Log$_{10}$ of M$_{\rm BH}$ (solar units). Random errors are estimated to be 0.4 dex. Systematic errors are discussed extensively in Section~\ref{sec:sys}}
\tablenotetext{a}{Spheroid size and luminosity are upper limits.}
\end{deluxetable}

\begin{deluxetable}{lrcc}
\tablewidth{0pt}
\tablecaption{Properties of the local comparison sample.}
\tablehead{
\colhead{Name}        &
\colhead{z}     &
\colhead{$\log L_{\rm B}/L_{\rm \odot, B}$}           &
\colhead{$\log M_{\rm BH} / M_{\rm \odot}$} 
\\
\colhead{(1)} &
\colhead{(2)} &
\colhead{(3)} &
\colhead{(4)}}
\tablecolumns{4}
\startdata
Ark120  & 0.032 & 10.82 & 8.18 \\
Mrk79   & 0.022 &  9.79 & 7.72 \\
Mrk110  & 0.035 &  9.85 & 7.40 \\
Mrk590  & 0.026 & 10.40 & 7.68 \\
Mrk817  & 0.031 & 10.49 & 7.69 \\
NGC3227 & 0.004 &  8.85 & 7.62 \\
NGC4051 & 0.002 &  8.43 & 6.28 \\
NGC4151 & 0.003 &  9.49 & 7.66 \\
NGC5548 & 0.017 & 10.53 & 7.83 \\
\enddata
\label{tab:local}
\tablecomments{
Col. (1): Target ID.  
Col. (2): Redshift.
Col. (3): Log$_{10}$ of spheroid Luminosity in rest frame B (solar units).
Col. (4): Log$_{10}$ of M$_{\rm BH}$ (solar units). From \citep{Pet++04} and \citep{Ben++06a}. Errors are estimated to be 0.4 dex.}
\end{deluxetable}

\clearpage

\end{document}